\documentclass[11pt,reqno]{article}

\usepackage{amsthm}
\usepackage[english]{babel}
\usepackage{amsmath}
\usepackage{amssymb}
\usepackage{graphicx}
\usepackage{color}
\usepackage{bm}
\usepackage{mathtools}
\usepackage{subfigure}

\newcommand{\nm}{\noalign{\smallskip}}
\newcommand{\ds}{\displaystyle}
\usepackage{pgf}
\usepackage{pgffor}
\usepgflibrary{plothandlers}
\usepackage{pgfplots}
\pgfplotsset{compat=newest}
 \pgfplotsset{width=15cm}
\pgfplotsset{plot coordinates/math parser=false}
\newlength\figureheight
\newlength\figurewidth

\newtheorem{theo}{Theorem}[section]

\newtheorem{prop}[theo]{Proposition}
\newtheorem{lemm}[theo]{Lemma}

\newcommand{\R}{\mathbb{R}}

\newcommand{\psibf}{{\psi}}

\newcommand{\phibf}{{\phi}}

\def\nm{\noalign{\medskip}}

\newcommand{\nubf}{{\nu}}

\numberwithin{equation}{section} \numberwithin{figure}{section}

\newcommand{\dr}{\partial}

\newcommand{\g}{\nabla}

\newcommand{\ol}{\overline}

\newcommand{\bE}{{E}}
\newcommand{\bH}{{H}}

\newcommand{\q}{\quad}

\def\nm{\noalign{\medskip}}

\newcommand{\beq}{\begin{equation}}
\newcommand{\eeq}{\end{equation}}

\DeclareMathAlphabet{\itbf}{OML}{cmm}{b}{it}

\begin{document}
\title{Surface Plasmon Resonance of Nanoparticles and Applications in Imaging\thanks{\footnotesize
This work was supported by the ERC Advanced Grant Project
MULTIMOD--267184.}}
\author{Habib Ammari\thanks{\footnotesize Department of Mathematics and Applications,
Ecole Normale Sup\'erieure, 45 Rue d'Ulm, 75005 Paris, France
(habib.ammari@ens.fr, deng@dma.ens.fr, pierre.millien@ens.fr).}
\and Youjun Deng\footnotemark[2] \and Pierre Millien\footnotemark[2]}
\maketitle

\begin{abstract}
In this paper we provide a mathematical framework for localized plasmon resonance of nanoparticles.  Using layer potential techniques associated with the full Maxwell equations, we derive small-volume expansions for the electromagnetic fields, which are uniformly valid with respect to the nanoparticle's bulk electron relaxation rate. Then, we discuss the scattering and absorption enhancements by plasmon resonant nanoparticles. We study both the cases of a single and multiple nanoparticles. We present numerical simulations of the localized surface plasmonic resonances associated to multiple particles in terms of their separation distance. \end{abstract}

\noindent {\footnotesize {\bf Mathematics subject classification
(MSC2000):} 35R30, 35C20}


%%%%%%%%%%%%%%%%%%%%%%%%%%%%%%%%%%%%%%%%%%%%%%%%%%%%%%
\section{Introduction}

Localized surface plasmons are charge density oscillations confined to metallic nanoparticles. 
Excitation of localized surface plasmons by an electromagnetic field at an incident wavelength where resonance occurs results in a strong light scattering and an enhancement of the local electromagnetic fields. Recently, the localized surface plasmon resonances of nanoparticles have received considerable attention for their application in biomedicine. They have enabled applications  including sensing of cancer cells and their photothermal ablation.  Plasmon resonant nanoparticles such as gold nanoparticles offer, in addition to their  enhanced scattering and absorption, biocompatibility making them  not only suitable for use as a contrast agent but also in therapeutic applications \cite{plasmon4}.

According to the quasi-static approximation for small particles, the surface plasmon resonance peak occurs when the particle's polarizability is maximized. Recently, it has been shown that plasmon resonances in nanoparticles can be treated as an eigenvalue problem for the Neumann-Poincar\'e operator, which leads to direct calculation of resonance values of permittivity and optimal design of nanoparticles that resonate at specified frequencies \cite{yatin, Gri12, plasmon3}. Classically, the frequency-dependent permittivity of metallic nanoparticles can be described by a Drude model which determines the material's dielectric and magnetic responses by considering the motion of the free electrons against a background of positive ion cores.

In this paper, we provide a rigorous mathematical framework for localized surface plasmon resonances. We consider the full Maxwell equations. Using layer potential techniques, we derive the quasi-static limits of the electromagnetic fields in the presence of  nanoparticles. We prove that the quasi-static limits 
 are uniformly valid with respect to the nanoparticle's bulk electron relaxation rate.
Note that uniform validity with respect to the contrast was proved in \cite{hoai} in the context of small volume expansions for the conductivity problem.   
  Then, we discuss the scattering and absorption enhancements by plasmon resonant nanoparticles. The nanoscale light concentration and near-field enhancement available to resonant metallic nanoparticles have been a driving force in nanoplasmonics. We first consider a single nanoparticle. Then we extend our approach to multiple nanoparticles. 
 We study the influence of local environment on the near-field behavior of resonant nanoparticles. We simulate the localized surface plasmonic resonances associated to multiple particles in terms of their separation distance.   

The paper is organized as follows. In section \ref{sec1}, we introduce  localized plasmonic resonances as the eigenvalues of the Neumann-Poincar\'e operator associated with the nanoparticle.   In section \ref{sec12} we describe a general model for the permittivity and permeability of nanoparticles as functions of the frequency. In section \ref{sec2}, we recall useful results on layer potential techniques for Maxwell's equations. Section \ref{sec22} is devoted to the derivation of the uniform asymptotic expansions. We rigorously justify the quasi-static approximation for surface plasmon resonances. Our main results are stated in Theorems \ref{theo2f} and \ref{theo3f}. In section \ref{sec23} we illustrate the validity of our results by a variety of numerical simulations. The paper ends with a short discussion.

\section{Plasmonic resonances} \label{sec1}
We first introduce the Neumann-Poincar\'e operator of an open connected domain $D$ with $\mathcal{C}^{1,\eta}$ boundary in $\mathbb{R}^d\;(d=2,3)$ for some $0<\eta<1$.  Given such a domain $D$, we consider the following Neumann problem,
\begin{equation}
        \Delta u = 0  \q \mbox{ in } ~~D\,; \q \q
        \frac{\partial u}{\partial \nu} = g  \q \mbox{ on } ~~\partial D,
        \q\q \int_{\partial D} u \, d\sigma =0, \label{neumann}
\end{equation}
where $g \in L^2_0(\partial D)$ with $L^2_0(\partial D)$ being the set of functions in $L^2(\partial D)$ with zero mean-value. In (\ref{neumann}),   $\partial/\partial \nu$ denotes the normal derivative.
We note that the Neumann problem (\ref{neumann}) can be rewritten as a boundary integral equation with the help of the single-layer potential. Given a density function $\varphi \in L^2(\partial D)$, the single-layer potential, $\mathcal{S}_{D} [\varphi]$, can be defined as follows,
\begin{equation}
    \mathcal{S}_{D} [\varphi] (x) := \int_{\partial D} \Gamma(x-y) \varphi(y) d \sigma(y)
\end{equation}
for $x \in \mathbb{R}^d$, where $\Gamma$ is the fundamental solution of the Laplacian in $\mathbb{R}^d$ :
\begin{equation}
    \Gamma (x-y) =
    \begin{cases}
     \frac{1}{2\pi} \log |x-y| & \mbox{ if }\; d = 2 \, ,\\
     \frac{1}{(2-d)\omega_d} |x-y|^{2-d} & \mbox{ if }\; d > 2 \, ,
    \end{cases}
    \label{fundamental}
\end{equation}
where $\omega_d$ denotes the surface area of the unit sphere in $\mathbb{R}^d$.  It is well-known that the single-layer potential satisfies the following jump condition on $\partial D$:
\begin{equation}
    \frac{\dr}{\dr \nu} \left( \mathcal{S}_{D} [\varphi] \right)^{\pm} = (\pm \frac{1}{2} I + \mathcal{K}^*_{  D} )[\varphi]\,,
    \label{jump_condition}
\end{equation}
where the superscripts $\pm$ indicate the limits from outside and inside $D$ respectively, and 
$\mathcal{K}^*_{D}: L^2(\partial D) \rightarrow L^2(\partial D)$ is the Neumann-Poincar\'e operator defined by
\begin{equation}
    \mathcal{K}^*_{ D} [\varphi] (x) := \frac{1}{\omega_d} \int_{\partial D} \frac{( x-y)\cdot \nu(x)  }{|x-y|^d} \varphi(y) d \sigma(y) 
    \label{operatorK}
\end{equation}
with $\nu(x)$ being 
the outward normal at $x \in \partial D$. We note that $\mathcal{K}^*_{D}$ maps $L^2_0(\partial D)$ onto itself.

With these notions, the Neumann problem (\ref{neumann}) can then be formulated as
\begin{equation}
    g = \left(\frac{\dr}{\dr \nu} ( \mathcal{S}_{D} [\varphi] \right)^{-} =  ( - \frac{1}{2} I + \mathcal{K}^*_{D} )[\varphi]  \, .
\end{equation}
Therefore, the solution to the Neumann problem (\ref{neumann}) can be reformulated as a solution to the boundary integral equation with the Neumann-Poincar\'e operator $\mathcal{K}^*_{D}$.

The operator $\mathcal{K}^*_{D}$ arises not only in solving the Neumann problem for the Laplacian but also for representing the solution to the transmission problem as described below.

Consider an open connected domain $D$ with $\mathcal{C}^2$ boundary in $\mathbb{R}^d$.  Given a harmonic function $u_0$ in $\mathbb{R}^d$, we consider the following transmission problem in $\mathbb{R}^d$:
\begin{equation}
    \begin{cases}
        \nabla \cdot (\varepsilon_{D} \nabla u) = 0 &\mbox{ in }\; \mathbb{R}^d, \\[1.5mm]
         u - u_0 = O(|x|^{1-d}) &\mbox{ as }\; |x| \rightarrow \infty,
    \end{cases}
    \label{transmission}
\end{equation}
where $\varepsilon_{D}  = \varepsilon_c \chi(D) +  \varepsilon_m \chi(\mathbb{R}^d \backslash \overline{D})$ with $\varepsilon_c, \varepsilon_m$ being two positive constants, and $\chi(\Omega)$ is the characteristic function of the domain $\Omega= D$ or $\mathbb{R}^d \backslash \overline{D}$.  With the help of the single-layer potential, we can rewrite the perturbation  $u - u_0$, which is  due to the inclusion $D$, as
\begin{equation}
    u - u_0 = \mathcal{S}_{D} [\varphi] \, ,
    \label{scattered}
\end{equation}
where $\varphi \in L^2(\partial D)$ is an unknown density, and $\mathcal{S}_{D} [\varphi]$ is the refraction part of the potential in the presence of the inclusion. The transmission problem (\ref{transmission}) can be rewritten as
\begin{equation}
    \begin{cases}
        \Delta u = 0 &\mbox{ in }\; D \bigcup (\mathbb{R}^d\backslash \overline{D} ) \, ,\\[1.5mm]
        u^+ = u^- &\mbox{ on }\; \partial D \, ,\\[1.5mm]
        \varepsilon_c \frac{\dr u^{+}}{\dr\nu}  =  \varepsilon_m \frac{\dr u^{-}}{\dr\nu} &\mbox{ on }\; \partial D \, ,\\[1.5mm]
        u - u_0 = O(|x|^{1-d}) &\mbox{ as }\; |x| \rightarrow \infty \, .
    \end{cases}
    \label{transmission2}
\end{equation}
With the help of the jump condition (\ref{jump_condition}), solving the above system (\ref{transmission2})
can be regarded as solving the density function $\varphi \in L^2(\partial D)$ of the following integral equation
\begin{equation}
    \frac{\partial u_0}{\partial \nu} =  \left( \frac{\varepsilon_c+\varepsilon_m}{2(\varepsilon_c-\varepsilon_m)}I - \mathcal{K}_{D}^* \right) [\varphi] \, .
    \label{potential2}
\end{equation}
With the harmonic property of $u_0$, we can write
\begin{equation}
    u_0 (x) = \sum_{\alpha \in \mathbb{N}^d } \frac{1}{\alpha !} \partial^\alpha u_0(0) x^\alpha
\label{series}
\end{equation}
with $\alpha=(\alpha_1,\ldots, \alpha_d) \in \mathbb{N}^d,\, \partial_\alpha = \partial_1^{\alpha_1} \ldots \partial_d^{\alpha_d}$ and $\alpha!=\alpha_1!\ldots \alpha_d!$\,.

Consider $\varphi^\alpha$ as the solution of the Neumann-Poincar\'e operator:
\begin{equation}
   \frac{\partial x^\alpha}{\partial \nu} =  \left( \frac{\varepsilon_c+\varepsilon_m}{2(\varepsilon_c-\varepsilon_m)}I - \mathcal{K}_{D}^* \right) [\varphi^\alpha] \, .
    \label{phi_alpha}
\end{equation}
The invertibilities of the operator $( \frac{\varepsilon_c+\varepsilon_m}{2(\varepsilon_c-\varepsilon_m)}I - \mathcal{K}_{D}^*)$
from $L^2(\partial D)$ onto $L^2(\partial D)$ and from
$L_0^2(\partial D)$ onto $L_0^2(\partial D)$ are proved, for example, in \cite{book2, kellog},
provided that $|\frac{\varepsilon_c+\varepsilon_m}{2(\varepsilon_c-\varepsilon_m)}| > 1/2$.
We can substitute (\ref{series}) and (\ref{phi_alpha}) back into (\ref{scattered}) to get
\begin{equation}
    u - u_0 = \sum_{|\alpha| \geq 1 }  \frac{1}{\alpha !} \partial^\alpha u_0(0) \mathcal{S}_{D} [\varphi^\alpha] = \sum_{|\alpha| \geq 1 } \frac{1}{\alpha !} \partial^\alpha u_0(0) \int_{\partial D} \Gamma(x-y) \varphi^\alpha (y) d \sigma(y) \, .
    \label{scattered2}
\end{equation}
Using the Taylor expansion,
\begin{equation}
\Gamma (x-y) = \Gamma(x) - y \cdot \nabla \Gamma(x) + O(\frac{1}{|x|^{d}})\, ,
\label{series2}
\end{equation}
which holds for all $x$ such that $|x|\rightarrow \infty$ while $y$ is bounded \cite{book2},
we get the following result by substituting (\ref{series2}) into (\ref{scattered2}) that
\begin{equation} \label{eq15}
    (u - u_0)(x) = \nabla u_0(0) \cdot M(\lambda, D) \nabla \Gamma (x) + O(\frac{1}{|x|^{d}}) \quad \mbox{as} ~~|x| \rightarrow \infty,
\end{equation}
where $M=(m_{ij})_{i,j=1}^d$ is the polarization tensor  associated with the domain $D$ and the contrast $\lambda$ defined by
\begin{equation}
    m_{ij}(\lambda,D) :=  \int_{\partial D} y_i (\lambda I - \mathcal{K}_{D}^*)^{-1} \left[\nu_j\right] (y) d \sigma(y) \, ,
    \label{polarizationtensionr}
\end{equation}
with 
\begin{equation} \label{deflambda} \lambda :=  \frac{\varepsilon_c+\varepsilon_m}{2(\varepsilon_c-\varepsilon_m)} 
\end{equation}  
and $\nu_j$ being the $j$-th component of $\nu$. Here we have used in (\ref{eq15}) the fact that
$\int_{\partial D} \nu \, d\sigma=0$.

Typically the constants $\varepsilon_c$ and $\varepsilon_m$ are positive in order to make the system (\ref{transmission2}) physical. This corresponds to the situation with $|\lambda| > \frac{1}{2}$.

However, recent advances in nanotechnology make it possible to produce noble metal nanoparticles
with negative permittivities at optical frequencies \cite{plasmon4, SC10}. 
Therefore, it is possible that for some frequencies, $\lambda$ actually belongs to the spectrum of $ \mathcal{K}_{D}^*$.

If this happens, the following integral equation
\begin{equation}
    0 =  \left( \lambda I - \mathcal{K}_{D}^* \right) [\varphi]   \quad \mbox{ on }\; \partial D \label{plasmonicreso}
\end{equation}
has non-trivial solutions $\varphi \in L^2(\partial D)$ and the nanoparticle resonates at those frequencies.

Therefore, we have to investigate the mapping properties of the Neumann-Poincar\'e operator. Assume that  $\partial D$ is of class $\mathcal{C}^{1,\eta}$, $0<\eta<1$. It is known that the operator $\mathcal{K}_{D}^*: L^2(\partial D) \rightarrow L^2(\partial D)$ is compact \cite{kellog}, and its spectrum is discrete and accumulates at zero. All the eigenvalues are real and bounded by $1/2$. Moreover, $1/2$ is always an eigenvalue and its associated eigenspace is of dimension one,
which is nothing else but the kernel of the single-layer potential $\mathcal{S}_{D}$.
In two dimensions, it can be proved that if $\lambda_i\neq 1/2$ is an eigenvalue of $\mathcal{K}_{D}^*$, then $-\lambda_i$ is an eigenvalue as well. This property is known as the twin spectrum property; see \cite{plasmon1}. The Fredholm eigenvalues are the eigenvalues of $\mathcal{K}_{D}^*$. It is easy to see, from the properties of $\mathcal{K}_{D}^*$, that they are invariant with respect to rigid motions and scaling.
They can be explicitly computed for ellipses and spheres.  If $a$ and $b$ denote the semi-axis lengths of an ellipse then it can be shown that $\pm (({a-b})/({a+b}))^i$ are its Fredholm eigenvalues \cite{shapiro}. For the sphere, they are given by $1/(2(2i+1))$; see \cite{seo}. It is worth noticing that the convergence to zero of Fredholm eigenvalues is exponential for ellipses while it is algebraic for spheres.

Equation (\ref{plasmonicreso}) corresponds to the case when plasmonic resonance occurs in $D$; see \cite{Gri12}.
Given negative values of $\varepsilon_c$, the problem of  designing a shape with prescribed plasmonic resonances is of great interest \cite{yatin}.

Finally, we briefly investigate the eigenvalue of the Neumann-Poincar\'e operator of 
multiple particles. Let $D_1$ and $D_2$ be two smooth bounded domains such that the distance $\mathrm{dist}(D_1,D_2)$ between $D_1$ and $D_2$ is positive. Let $\nu^{(1)}$ and $\nu^{(2)}$ denote the outward normal vectors at $\partial D_1$ and $\partial D_2$, respectively. 
 
The Neumann-Poincar\'e operator $\mathbb{K}^*_{D_1 \cup D_2}$ associated with $D_1 \cup D_2$ is given by \cite{ciraolo}
\begin{eqnarray}
     \mathbb{K}^*_{D_1 \cup D_2}:=
    \begin{pmatrix}
    \mathcal{K}^*_{D_1} & \frac{\partial}{\partial \nu^{(1)}} \mathcal{S}_{ D_2}\\
    \frac{\partial}{\partial \nu^{(2)}} \mathcal{S}_{D_1} &\mathcal{K}^*_{D_2}
    \end{pmatrix} .
    \label{matrixda}
\end{eqnarray}
In section \ref{sec23} we will be interested in how the eigenvalues of $\mathbb{K}^*_{D_1 \cup D_2}$ behave numerically as $\mathrm{dist}(D_1,D_2)\rightarrow 0$.

\section{Drude's model for the electric permittivity and magnetic permeability} \label{sec12}
Let $D$ be a bounded domain in $\mathbb{R}^d$ with
$\mathcal{C}^{1, \eta}$ boundary for some $0<\eta<1$, and let
$(\varepsilon_m,\mu_m)$ be the pair
 of electromagnetic parameters (electric permittivity and magnetic permeability) of $\mathbb{R}^d \setminus \overline{D}$
 and $(\varepsilon_c, \mu_c)$ be that of $D$. We assume that $\varepsilon_m$ and $\mu_m$ are real positive constants. We have
\begin{equation*}
\varepsilon_D=\varepsilon_m \chi ( \mathbb{R}^d \setminus \overline{D}) +
\varepsilon_c \chi(D) \quad \mbox{and} \quad \mu_D =\mu_m \chi ( \mathbb{R}^d
\setminus \overline{D}) + \mu_c \chi(D).
\end{equation*}

Suppose that the electric permittivity $\varepsilon_c$ and the magnetic permeability $\mu_c$ of the nanoparticle are changing with respect to the operating angular frequency $\omega$ while those of the surrounding medium, $\varepsilon_m, \mu_m$, are independent of $\omega$. Then we can write
\begin{equation}
\begin{aligned}
\varepsilon_c(\omega)= \varepsilon'(\omega) + i \varepsilon''(\omega),\\
\mu_c(\omega)= \mu'(\omega) + i \mu''(\omega).
\end{aligned}
\end{equation}
Because of causality, the real and imaginary parts of $\varepsilon_c$ and $\mu_c$ obey the following Kramer--Kronig relations: 
\begin{equation}
\begin{aligned}
&\varepsilon'(\omega)= -\frac{1}{\pi} \mathrm{p.v. } \int_{-\infty}^{+\infty}\frac{1}{\omega-s}\varepsilon''(s)ds,\\
&\varepsilon''(\omega)= \frac{1}{\pi} \mathrm{p.v. } \int_{-\infty}^{+\infty}\frac{1}{\omega-s}\varepsilon'(s)ds,\\
&\mu''(\omega)=- \frac{1}{\pi} \mathrm{p.v. } \int_{-\infty}^{+\infty}\frac{1}{\omega-s}\mu'(s)ds,\\
&\mu'(\omega)= \frac{1}{\pi} \mathrm{p.v. } \int_{-\infty}^{+\infty}\frac{1}{\omega-s}\mu''(s)ds,
\end{aligned}
\end{equation} 
where $\mathrm{p.v.}$ denotes the principle value. 

In the sequel, we set $k_c=\omega \sqrt{\varepsilon_c \mu_c}$ and $k_m=\omega \sqrt{\varepsilon_m \mu_m}$ and denote by 
\begin{equation} \label{deflambdaomega} \lambda_\varepsilon(\omega)= \frac{\varepsilon_c(\omega)+\varepsilon_m}{2(\varepsilon_c(\omega)-\varepsilon_m)}, \quad 
\lambda_\mu (\omega) = \frac{\mu_c(\omega)+\mu_m}{2(\mu_c(\omega)-\mu_m)}.
\end{equation}
We have 
$$
\lambda_\varepsilon(\omega)= \frac{(\varepsilon^\prime(\omega))^2 -\varepsilon_m^2 + 
(\varepsilon^{\prime\prime}(\omega))^2}{2 ( (\varepsilon^\prime(\omega) - \varepsilon_m)^2 + 
 (\varepsilon^{\prime\prime}(\omega))^2} - i \frac{\varepsilon^\prime(\omega) \varepsilon^{\prime\prime}(\omega)}{2 ( (\varepsilon^\prime(\omega) - \varepsilon_m)^2 + 
 (\varepsilon^{\prime\prime}(\omega))^2}.$$
 A similar formula holds for $\lambda_\mu(\omega)$.

The electric permittivity $\varepsilon_c(\omega)$ and the magnetic permeability $\mu_c(\omega)$  can be described by the Drude Model; see, for instance, \cite{SC10}. We have
$$
\varepsilon_c(\omega)=\varepsilon_0(1-\frac{\omega_p^2}{\omega(\omega+i\tau^{-1})}) \quad \mbox{and} \quad \mu_c(\omega)=\mu_0(1-F\frac{\omega^2}{\omega^2-\omega_0^2+i\tau^{-1} \omega}),
$$
or equivalently, 
$$\begin{array}{l}
\ds
\varepsilon^\prime(\omega)= \varepsilon_0 \frac{\omega^2 + \tau^{-2} - \omega_p^2}{\omega^2 + \tau^{-2}}, \quad \varepsilon^{\prime\prime}(\omega) = \varepsilon_0 \frac{\omega_p^2 \tau^{-1}}{\omega( \omega^2 + \tau^{-2})},\\
\nm
\ds
\mu^\prime(\omega)= \frac{\mu_0 (\tau^{-2} \omega^2 + (\omega^2 - \omega_0^2)((1-F) \omega^2 - \omega_0^2)}{(\omega^2 - \omega_0^2)^2 + \tau^{-2}\omega^2}, \quad 
\mu^{\prime\prime}(\omega) = \frac{\mu_0 F \tau^{-1} \omega}
{(\omega^2 - \omega_0^2)^2 + \tau^{-2} \omega^2 },
\end{array}$$
where $\omega_p$ is the plasma frequency of the bulk
material, $\tau >0$ is the nanoparticle's bulk electron relaxation rate ($\tau^{-1}$ is the damping coefficient), $F$ is a filling factor, and $\omega_0$ is a localized plasmon resonant frequency. 

When
$$\omega^2 + \tau^{-2} <\omega_p^2  \quad \mbox{ and } (1-F) (\omega^2 - \omega_0^2)^2 - F \omega_0^2 (\omega^2 - \omega_0^2) + \tau^{-2} \omega^2 < 0, $$ the real parts of $\varepsilon(\omega)$ and $\mu(\omega)$ are negative. Typical values are 
\begin{itemize}
\item $\tau = 10^{-14}\, s$; 
\item $\omega = 10^{15}\, Hz$;
\item $\varepsilon_0 = 9 \cdot 10^{-12} F \, m^{-1}$; $\varepsilon_m = (1.33)^2 \varepsilon_0$;
\item $\omega_p = 2 \cdot 10^{15} s^{-1}$ for a gold nanoparticle;
%\item $\varepsilon'_c(10^{15}) \approx -2\cdot  10^{-9}$ and $\varepsilon_c''(10^{15}) %\approx 2 \cdot 10^{-9}$ for a gold nanoparticle.
% $\lambda_\varepsilon =0.495 - 0.005i $ si $\varepsilon_m = 2*\varepsilon_0$. 
\end{itemize}
%Using these values we find that $\lambda_\varepsilon \approx 0.495 - 0.005$. 

It is interesting to have an idea on the size of $\Im m(\lambda_\varepsilon)$  
(resp. $\Im m(\lambda_\mu)$)
since it will be a lower bound for the distance  $\mathrm{dist}(\lambda_\varepsilon, \sigma(\mathcal{K}^*_D))$ (resp. $\mathrm{dist}(\lambda_\mu, \sigma(\mathcal{K}^*_D))$) between $\lambda_\varepsilon$ (resp. $\lambda_\mu$) and the spectrum of the Neumann-Poincar\'e operator $\mathcal{K}_D^*$.

Finally, we define dielectric and magnetic plasmonic resonances. We say that $\omega$ is a dielectric plasmonic resonance if the real part of 
$\lambda_\varepsilon$ is an eigenvalue  of $ \mathcal{K}_{D}^*$. Analogously, we say that  $\omega$ is a magnetic plasmonic resonance if the real part of 
$\lambda_\mu$ is an eigenvalue  of $ \mathcal{K}_{D}^*$.
Note that if $\omega$ is a dielectric (resp. magnetic) plasmonic resonance,  then the polarization tensor 
$M(\lambda_\varepsilon(\omega), D)$ defined by (\ref{polarizationtensionr}) (resp. $M(\lambda_\mu(\omega), D)$) blows up.

In the case of two particles $D_1$ and $D_2$ with the same electromagnetic parameters,
$\varepsilon_c(\omega)$ and $\mu_c(\omega)$, we say that $\omega$ is a dielectric (resp. magnetic) plasmonic resonance,  if the real part of 
$\lambda_\varepsilon$ is an eigenvalue  of $ \mathbb{K}_{D_1 \cup D_2}^*$.  Analogously, we say that $\omega$ is a magnetic plasmonic resonance if the real part of 
$\lambda_\mu$ is an eigenvalue  of
 $ \mathbb{K}_{D_1 \cup D_2}^*$.

Let the polarization tensor 
$M(\lambda, D_1 \cup D_2)=(m_{ij})_{i,j=1}^d$ be defined by 
\begin{equation}
\begin{array}{lll}
    m_{ij}(\lambda,D_1 \cup D_2) &:=&\ds  \int_{\partial D_1} y_i \bigg[ (\lambda I - \mathbb{K}_{D}^*)^{-1} \left[ \begin{matrix} \nu_j^{(1)} \\ \nu_j^{(2)} \end{matrix} \right] (y) \bigg]_1 \, d \sigma(y) \\ \nm  && \ds + \int_{\partial D_2} y_i \bigg[ (\lambda I - \mathbb{K}_{D}^*)^{-1} \left[ \begin{matrix} \nu_j^{(1)} \\ \nu_j^{(2)} \end{matrix} \right] (y) \bigg]_2 \, d \sigma(y) \, ,
    \end{array}
    \label{polarizationtensionrm}
\end{equation}
where $\nu^{(l)}=(\nu^{(l)}_1, \ldots, \nu^{(l)}_d)$, $l=1,2$, and  $[\,]_{l^\prime}$ denotes the $l^\prime$th component. As for single particles, $M(\lambda(\omega), D_1 \cup D_2)=(m_{ij})_{i,j=1}^d$ blows up for $\lambda(\omega)$ such that $\omega$ is a dielectric or magnetic plasmonic resonance.

\section{Boundary integral operators}\label{sec2}

We start by recalling some well-known properties about boundary integral operators and proving a few technical lemmas that will be used in section \ref{sec22} for deriving the asymptotic expansions of the electric and magnetic fields in the presence of nanoparticles. As will be shown in section \ref{sec23}, the plasmonic resonances for multiple identical particles are shifted from those of the single particle as the separating distance between the particles becomes comparable to their size.

\subsection{Definitions}

We first review commonly used function spaces. 
 Let  $\nabla_{\dr D}\cdot$ denote the surface divergence. Denote by $L_T^2(\dr D):=\{\phibf\in {L^2(\dr D)}^3, \nubf\cdot \phibf=0\}$. Let $H^s(\partial D)$ be the usual Sobolev space of order $s$ on $\partial D$. We also introduce the function
spaces
\begin{align*}
\mathrm{TH}({\rm div}, \dr D):&=\Bigr\{ {\phi} \in L_T^2(\partial D):
\nabla_{\partial D}\cdot {\phi} \in L^2(\partial D) \Bigr\},\\
\mathrm{TH}({\rm curl}, \dr D):&=\Bigr\{ {\phi} \in L_T^2(\partial D):
\nabla_{\partial D}\cdot ({\phi}\times {\nu}) \in L^2(\partial D) \Bigr\},
\end{align*}
equipped with the norms
\begin{align*}
&\|{\phi}\|_{\mathrm{TH}({\rm div}, \dr D)}=\|{\phi}\|_{L^2(\dr D)}+\|\nabla_{\dr D}\cdot {\phi}\|_{L^2(\dr D)}, \\
&\|{\phi}\|_{\mathrm{TH}({\rm curl}, \dr D)}=\|{\phi}\|_{L^2(\dr D)}+\|\nabla_{\dr D}\cdot({\phi}\times \nu)\|_{L^2(\dr D)}.
\end{align*}
We define the vectorial curl for $\varphi \in H^1(\partial D)$ by $\mathrm{curl}_{\partial D} \varphi = - \nu \times \nabla_{\partial D} \varphi$.

The following result from \cite{buffa2002traces} will be useful.
\begin{prop} The following Helmholtz decomposition holds:
\begin{equation} \label{helmdecom}
\mathrm{L}^2_T(\partial D) = \nabla_{\partial D} (H^1(\partial D)) \overset{\perp}{\oplus}  \mathrm{curl}_{\partial D} (H^1(\partial D) ).
\end{equation}
\end{prop}

Next, we recall that, for $k>0$, the fundamental outgoing solution $\Gamma^k$ to the Helmholtz
operator $(\Delta+k^2)$ in $\mathbb{R}^3$ is given by
\begin{equation}\label{Gk} \ds\Gamma^k
(x) = -\frac{e^{ik|x|}}{4 \pi |x|}.
 \end{equation}

For a density $\phibf \in \mathrm{TH}({\rm div}, \dr D)$, we define the
vectorial single layer potential associated with the fundamental solution
$\Gamma^k$ introduced in (\ref{Gk}) by
\begin{equation}\label{defA}
\ds\mathcal{A}_D^{k}[\phibf](x) := \int_{\dr D} \Gamma^k(x-y)
\phibf(y) d \sigma(y), \quad x \in \mathbb{R}^3.
\end{equation}
For a scalar density $\varphi \in L^2(\dr D)$, the single layer
potential is defined similarly by
\begin{equation}\label{defS}
\mathcal{S}_D^{k}[\varphi](x) := \int_{\dr D} \Gamma^k(x-y) \varphi(y) d \sigma(y), \quad x \in \mathbb{R}^3.
\end{equation}
We will also  need the following boundary operators:
\begin{equation}\label{defM}
\begin{aligned} \mathcal{M}_D^k: 
\mathrm{L}_T^2 (\partial D)  & \longrightarrow \mathrm{L}_T^2 (\partial D)  \\
\phibf & \longmapsto \mathcal{M}^k_D[\phibf]= \nubf(x)  \times \g \times \int_{\dr D} \Gamma^k(x,y) \nubf(y) \times \phibf(y) d\sigma(y),
\end{aligned}
\end{equation}
\begin{equation}\label{defN}
\begin{aligned} \mathcal{N}_D^k: 
\mathrm{TH}({\rm curl}, \dr D) & \longrightarrow \mathrm{TH}({\rm div}, \dr D) \\
\phibf & \longmapsto \mathcal{N}^k_D[\phibf]= 2 \nubf(x)  \times \g \times \g\times \int_{\dr D} \Gamma^k(x,y) \nubf(y) \times \phibf(y) d\sigma(y),
\end{aligned}
\end{equation}
\begin{equation}\label{defL}
\begin{aligned} \mathcal{L}_D^k: 
\mathrm{TH}({\rm div}, \dr D) & \longrightarrow \mathrm{TH}({\rm div}, \dr D) \\
\phibf & \longmapsto \mathcal{L}^k_D[\phibf]= \nubf(x)  \times k^2\mathcal{A}_D^k[\phibf](x) + \g\mathcal{S}_D^k[\g_{\dr D} \cdot \phibf](x).
\end{aligned}
\end{equation}

In the following, we denote by $\mathcal{A}_D$, $\mathcal{S}_D$, $\mathcal{M}_D$, and $\mathcal{N}_D$ the operators $\mathcal{A}_D^0$, $\mathcal{S}_D^0$, $\mathcal{M}_D^0$, and $\mathcal{N}_D^0$ corresponding to $k=0$, respectively.

\subsection{Boundary integral  identities}

Let $\mathcal{K}_D$ be the $L^2$-adjoint of $\mathcal{K}^*_D$ defined in (\ref{operatorK}). Since 
$\mathcal{K}_D$ and $$\mathcal{K}^*_D: L^2(\partial D) \rightarrow L^2(\partial D)$$ are compact and all the eigenvalues of $\mathcal{K}^*_D$ are real, we have $\sigma(\mathcal{K}_D) = \sigma(\mathcal{K}^*_D)$.

We start with stating the following jump formula. We refer the reader to  Appendix \ref{jumpMproof} for its proof.

\begin{prop}\label{propjumpM}

Let $ \phibf \in L^2_T(\dr D)$. Then $\mathcal{A}_D^k[\phibf]$ is continuous on $\mathbb{R}^3$ and its curl satisfies the following jump formula:

\begin{equation}\label{jumpM}
\left( \nubf \times \g \times \mathcal{A}_D^k[\phibf]\right)^\pm = \mp \frac{\phibf}{2} + \mathcal{M}_D^k[\phibf] \quad \mbox{ on } \dr D,
\end{equation}
where \begin{equation*}
\forall x\in \dr D, \quad \left( \nubf(x) \times \g \times \mathcal{A}_D^k[\phibf]\right)^\pm (x)= \lim_{t\rightarrow 0^+} \nubf(x) \times \g \times \mathcal{A}_D^k[\phibf] (x\pm t \nubf(x)).
\end{equation*} 
\end{prop}

Next, we prove the following integral identities.
\begin{prop}
We have 
\begin{equation}\label{defMstar}
\mathcal{M}_D^* = r\mathcal{M}_D r,
\end{equation}
where $r$ is defined by 
\begin{equation}\label{defr}
r[\phibf] = \nubf\times\phibf, \quad \forall \phibf \in L^2_T(\dr D).
\end{equation}
Moreover,
\begin{equation}\label{divA}
\g\cdot \mathcal{A}^k_{D}[\phibf] = \mathcal{S}_D^k[\g_{\dr D} \cdot \phibf] \quad \mbox{ in } \mathbb{R}^3, \quad \forall \phibf \in TH\left(\mathrm{div},\dr D\right).
\end{equation}
\begin{equation}\label{divM}
\g_{\dr D }\cdot\mathcal{M}^k_D [\phibf ] = -k^2 \nu \cdot \mathcal{A}_{ D}^k[\phibf] - \left(\mathcal{K}_{ D}^{k}\right)^* [\g_{\dr B }\cdot \mathrm{\phibf} ], \quad \forall \phibf \in TH\left(\mathrm{div},\dr D\right).
\end{equation}
Furthermore, 
 \begin{equation}\label{divM0}
\g_{\dr D }\cdot\mathcal{M}_D [\phibf ] =-\mathcal{K}_D^*[\g_{\dr D} \cdot \phibf], \quad \forall \phibf \in TH\left(\mathrm{div},\dr D\right),
\end{equation}
\begin{equation}\label{Metoilegrad}
\mathcal{M}^*_{D} [\nabla_{\partial D} \phi ] =- \nabla_{\partial D} \mathcal{K}_D [\phi],
\end{equation}
and 
\begin{equation}\label{curl}
\mathcal{M}_D [\mathrm{curl}_{\partial D} \phi ] =\mathrm{curl}_{\partial D}\mathcal{K}_D[\phi], \quad \forall \phibf \in TH(\mathrm{curl}, \partial D).
\end{equation}
\end{prop}
\begin{proof}
 The proof of (\ref{divA}) can be found in \cite{colton2012inverse}. We give it here for the sake of completeness. 
If  $\phibf \in TH\left(\mathrm{div},\dr D\right)$, then 
\begin{equation*}
\g \cdot \mathcal{A}^k_D[\phibf](x) = \int_{\dr D} \g_x \cdot \left( \Gamma^k(x,y) \phibf (y) \right) d\sigma(y), \quad x\in \mathbb{R}^3\setminus \dr D.
\end{equation*}
Using the fact that $$\g_x \cdot (\Gamma^k(x,y)\phibf(y)) = \phibf(y) \g_x \Gamma^k(x,y) = - \phibf(y) \g_y \Gamma^k(x,y),$$ and since $$\g_y \Gamma^k(x,y) = \g_{\dr D} \Gamma^k(x,y)  + \frac{\dr \Gamma^k}{\dr \nu (y)}(x,y),$$ we get
\begin{equation*}
\g \cdot \mathcal{A}^k_D[\phibf](x) = -\int_{\dr D} \phibf(y)\cdot \g_{\dr D,y} \Gamma^k(x,y)  d\sigma(y), \quad x\in \mathbb{R}^3\setminus \dr D.
\end{equation*}
Using the fact that $-\g_{\dr D}$ is the adjoint of $\g_{\dr_D} \cdot$ we obtain
\begin{equation*}
\g \cdot \mathcal{A}^k_D[\phibf](x) = \int_{\dr D} \Gamma^k(x,y)\g_{\dr D}\cdot \phibf(y)  d\sigma(y), \quad x\in \mathbb{R}^3\setminus \dr D.
\end{equation*}
Next, since $\mathcal{S}^k[\g_{\dr D}\phibf]$ is continuous across $\dr D$, the above relation can be extended to $\mathbb{R}^3$ and we get (\ref{divA}).

Now, in order to prove (\ref{divM}), we observe that, for any $\phibf \in TH\left(\mathrm{div},\dr D\right)$, 
\begin{equation*}
\g \times \g \times \mathcal{A}^k_D[\phibf](x) = k^2 \mathcal{A}^k_D[\phibf](x) + \g \mathcal{S}^k_D[\g_{\dr D}\cdot \phibf](x), \quad  x\in \mathbb{R}^3\setminus \dr D.
\end{equation*}
Using the jump relations on $\displaystyle{\frac{\dr \mathcal{S}^k_D}{\dr \nu}}$ we obtain that
\begin{equation*}
2 \left( \nu \cdot \g \times \g \times \mathcal{A}^k_D[\phibf] \right)^\pm = k^2\nu \cdot \mathcal{A}_D^k + (\mathcal{K}_D^k)^* [\g_{\dr D}\cdot \phibf] \mp  \g_{\dr D}\cdot \phibf \quad \mbox{ on } \dr D.
\end{equation*}
Recall from \cite[p.169]{colton2012inverse}  that if $\mathrm{f}\in \mathcal{C}^1(\mathbb{R}^3\setminus \overline{D})\cap \mathcal{C}^0(\mathbb{R}^3\setminus D)$, then $\g_{\dr D}\cdot (\nu \times \mathrm{f}) = -\nu\cdot (\g \times \mathrm{f}).$ Using the jump formula for $2 \left(\nu\times \g \times \mathcal{A}_D^k[\phibf]\right)^\pm = \mathcal{M}_D^k[\phibf] \pm \phibf$, we arrive at (\ref{divM}). Setting $k=0$ in (\ref{divM}) gives (\ref{divM0}).

Identity (\ref{Metoilegrad}) can be deduced from (\ref{divM0}) by duality.

Now, we prove (\ref{curl}). Define $r[{a}]= {\nu} \times {a}$ for any smooth vector field ${a}$ on $\partial D$. For $\phi \in H^1(\partial D)$, we have
\begin{equation*}
\mathcal{M}^*_{D} [\nabla_{\partial D} \phi ] =- \nabla_{\partial D} \mathcal{K}_D [\phi].
\end{equation*}
Since $\mathcal{M}_D^*=r \mathcal{M}_D r$ (see \cite{griesmaier2008asymptotic}) and $\mathrm{curl}_{\partial D} = r[\nabla_{\partial D}]$, it follows that
\begin{equation*}
r \left( \mathcal{M}_D [\mathrm{curl}_{\partial D} \phi ] \right)= -\nabla_{\partial D} \mathcal{K}_D [\phi].
\end{equation*}
Composing by $r^{-1}=-r$, we get
\begin{equation*}
\mathcal{M}_D [\mathrm{curl}_{\partial D} \phi ] = \mathrm{curl}_{\partial D} \mathcal{K}_ D [\phi],
\end{equation*}
which completes the proof. \end{proof}

\begin{lemm}\label{kernelN}
The kernel of the operator $\mathcal{N}_D$ in $L^2_T(\dr D)$ is $\g_{\dr D} (H^1(\dr D))$.
\end{lemm}
\begin{proof} Take $\phibf=\mathrm{curl}_{\dr D} U$ with $U \in H^1(\dr D)$. From (\ref{divA}), it follows that
\begin{equation*}
\mathcal{N}_D[\mathrm{curl}_{\dr D}U](x) = 2 \nubf(x)\times \g \mathcal{S}_D[ \g_{\dr D} \cdot \mathrm{curl}_{\dr D} U].
\end{equation*}
Since $\g_{\dr D} \cdot \mathrm{curl}_{\dr D} U=0$, we have $\mathcal{N}_D[\phibf]=0$.
Now, take $\phibf \in L^2_T(\dr D)$ such that $\mathcal{N}_D[\phibf]=0$. Then, 
on $\dr D$, we have
\begin{equation*}
\begin{aligned}
 2 \nubf(x)\times \g \mathcal{S}_D[ \g_{\dr D} \cdot\phibf]=& 2 \nubf(x)\times \left( \g_{\dr D} \mathcal{S}_D[ \g_{\dr D} \cdot r[\phibf]] + \frac{\dr}{\dr \nubf }\mathcal{S}_D[ \g_{\dr D} \cdot(r[\phibf]]\right)\\
 =& - 2 \mathrm{curl}_{\dr D}\mathcal{S}_D[ \mathrm{curl}_{\dr D}\phibf].
\end{aligned}
\end{equation*} Since  $\mathrm{Ker}(\mathrm{curl}_{\dr D})=\mathbb{R}$ (see \cite{buffa2002traces}), we obtain that $\mathcal{S}_{\dr D}[\mathrm{curl}_{\dr D}\phibf]=c\in\mathbb{R}$. Then, $\mathrm{curl}_{\dr D}\phibf =0$, which implies that $\phibf \in \g_{\dr D} H^1(\dr D)$ (see again \cite{buffa2002traces}).
\end{proof}

\begin{prop}We have the following Calder\'on type identity:
\begin{equation}\label{calderon}
\mathcal{N}_D \mathcal{M}^*_D= \mathcal{M}_D\mathcal{N}_D .
\end{equation}
\end{prop}
\begin{proof}
Let $\phibf \in H^{1/2}(\dr D)$. We have
$$\begin{array}{lll}
\ds \mathcal{M}_D \mathcal{N}_D [\phibf] &=&\ds  2\mathcal{M}_D\Big[ r\left( \g\times \g \times \mathcal{A}_D \big[r[\phibf] \big]\right) \Big],\\
\nm
& =&\ds 2\mathcal{M}_D\Big[ r\left(  \g\mathcal{S}_D \big[ \g_{\dr D} \cdot r[\phibf] \big]\right) \Big].
\end{array}$$
Since 
\begin{equation*}
\begin{aligned}
r\Big(\g \mathcal{S}_D[\g_{\dr D}\cdot r[\phibf]]\Big) = &\nubf \times \big[\g_{\dr D}  \mathcal{S}_D[\g_{\dr D}\cdot r[\phibf]] + \frac{\dr}{\dr \nubf}  \mathcal{S}_D[\g_{\dr D}\cdot r[\phibf]]\nubf\big]\\
=& - \mathrm{curl}_{\dr D} \mathcal{S}_D[\g_{\dr D}\cdot r[\phibf]],
\end{aligned}
\end{equation*} we can deduce from (\ref{curl}) that 
\begin{equation*}
\mathcal{M}_D\mathcal{N}_D [\phibf]  = - 2 \mathrm{curl}_{\dr D} \Big(\mathcal{K}_D \mathcal{S}_D \left[\g_{\dr D}\cdot r[\phibf]\right]\Big) .
\end{equation*}
Now, using the fact that $\mathcal{M}_D^*=r\mathcal{M}_D r$ and that $r^{-1}= - r$, we also have
$$\begin{array}{lll}
\mathcal{N}_D\mathcal{M}_D^* [\phibf] &=&\ds -2 r\left( \g\times \g \times \mathcal{A}_D \mathcal{M}_D\big[r[\phibf] \big]\right),
\\
\nm
\ds
&=&\ds -2r\Big( \g\mathcal{S}_D \left[ \g_{\dr D} \cdot \mathcal{M}_D\big[r[\phibf] \big]\right]\Big).
\end{array}$$
Moreover, (\ref{divM0}) yields 
\begin{equation*}
\mathcal{N}_D\mathcal{M}_D^* [\phibf] =2r\Big( \g\mathcal{S}_D \left[\mathcal{K}^*_D \big[\g_{\dr D}  \cdot r[\phibf] \big]\right]\Big).
\end{equation*}
Using Calder\'on's identity  $\mathcal{S}_B \mathcal{K}^*_B = \mathcal{K}_B \mathcal{S}_B$ and the fact that $$r[\g \mathcal{K}_D] = r[\g_{\dr D} \mathcal{K}_D ]= - \mathrm{curl}_{\dr D} \mathcal{K}_D,$$ it follows that
\begin{equation*}
\mathcal{N}_D\mathcal{M}_D^* [\phibf] = - 2\mathrm{curl}_{\dr D} \left(\mathcal{K}_D \mathcal{S}_D \big[\g_{\dr D}  \cdot r[\phibf] \big]\right),
\end{equation*}
which completes the proof. \end{proof}

\subsection{Resolvent estimates}
As seen in the section \ref{sec1}, we have to solve Fredholm type equations involving the resolvent of $\mathcal{K}_D$. We will also need to control the resolvent of $\mathcal{M}_D$. For doing so, the main difficulty is due to the fact that $\mathcal{K}_D$ and $\mathcal{M}_D$ are not self-adjoint. However, we will make use of a symmetrization technique in order to estimate the norms of the resolvents of $\mathcal{K}_D$ and $\mathcal{M}_D$.

The following result holds. 
\begin{prop}\label{resolvantK}
The operator 
$\mathcal{K}_D: L^2 (\partial D)   \longrightarrow  L^2(\partial D)$
satisfies the following resolvent estimate
\begin{equation*}
\Vert (\lambda I - \mathcal{K}_D)^{-1} \Vert_{\mathcal{L}(L^2(\partial D),L^2(\partial D)) } \leq \frac{C}{\mathrm{dist}(\lambda, \sigma(\mathcal{K}_D)) },
\end{equation*} where $\mathrm{dist}(\lambda, \sigma(\mathcal{K}_D))$ is the distance between $\lambda$ and  the spectrum $\sigma(\mathcal{K}_D)$ of $\mathcal{K}_D$, $C$ is a constant depending only on $D$, and $\mathcal{L}(L^2(\partial D),L^2(\partial D))$ is the set of bounded linear operators from $L^2(\partial D)$ into $L^2(\partial D)$.
\end{prop}
\begin{proof}
We start from Calder\'on's identity 
\begin{equation*}
\forall \phi \in L^2(\partial D), \quad \mathcal{S}_D \mathcal{K}^*_D [\phi] = \mathcal{K}_D \mathcal{S}_D [\phi].
\end{equation*}

Since $- \mathcal{S}_D: 
H^{-1/2} (\partial D)  \longrightarrow  H^{1/2}(\partial D)$ is a self-adjoint positive definite invertible operator in dimension three, we can define a new inner product on $H^{-1/2}(\partial D)$. We denote $\mathcal{H}$ the Hilbert space  $H^{-1/2}(\partial D)$ equipped with the  following inner product 
\begin{equation*}
\langle\phi,\psi\rangle_\mathcal{H} = - \langle \mathcal{S}_D [\phi] , \psi \rangle_{H^{1/2}, H^{-1/2}}  \quad \forall (\phi,\psi)\in  \left(H^{-1/2}(\partial D)\right) ^2
\end{equation*}
with $ \langle \; , \;\rangle_{H^{1/2}, H^{-1/2}}$ being the duality pairing between $H^{1/2}(\partial D)$ and $H^{-1/2}(\partial D)$.  
Now, $\sqrt{-\mathcal{S}_D}^{-1} \mathcal{K}_D \sqrt{-\mathcal{S}_D}$ is a self-adjoint compact operator on $\mathcal{H}$ and hence, we can write \cite{gilʹ1995norm}
\begin{equation*}
\Vert (\lambda I - \sqrt{-\mathcal{S}_D}^{-1} \mathcal{K}_D \sqrt{-\mathcal{S}_D})^{-1} \Vert_{\mathcal{L}(\mathcal{H}, \mathcal{H})} \leq \frac{C}{\mathrm{dist}(\lambda, \sigma(\mathcal{K}_D))}
\end{equation*}
for some constant $C$. Since $\sqrt{-\mathcal{S}_D}: H^{-1/2}(\partial D) \mapsto L^2(\partial D)$ is continuous and invertible, switching back to the original norm we get the desired result. \end{proof}

\begin{prop}\label{spectrumequality}
We have $\sigma(\mathcal{M}_D)= \big( -\sigma(\mathcal{K}_D^*)  \cup \sigma(\mathcal{K}_D^*) \big) 
\setminus\{\frac{1}{2} \}$.

\end{prop}
\begin{proof}
First, we note that $-1/2$ is not an eigenvalue of  $\mathcal{M}_D$; see \cite{griesmaier2008asymptotic, mitrea1996vector}. 
Let $\lambda \in \sigma(\mathcal{M}_D)$. Take $\phi\in {L}^2_T(\partial D) $ such that \begin{equation}\label{vecteurpropre}
\left(\lambda I - \mathcal{M} \right) [\phibf]= 0
\end{equation}
Using the Helmholtz decomposition (\ref{helmdecom}), we write 
\begin{equation*}
\phibf =\g_{\dr D} U + \mathrm{curl}_{\partial D} V.
\end{equation*}
Equation (\ref{vecteurpropre})  becomes 
\begin{equation}\label{vecteurspropres2}
\left(\lambda I - \mathcal{M}_D\right) \big[ \g_{\dr D} U + \mathrm{curl}_{\partial D} V\big] = 0,
\end{equation}
which yields
\begin{equation*}
\mathrm{curl}_{\partial D} \left(\lambda I - \mathcal{K}_D\right) [V]= - \left(\lambda I - \mathcal{M}_D\right) [\g_{\dr D} U ].
\end{equation*}
Taking the surface divergence we get
\begin{equation*}
\lambda \Delta_{\dr D} U - \g_{\dr D} \cdot \mathcal{M}_D [\g_{\dr D} U] = 0,
\end{equation*}
and hence, by using (\ref{divM0}), 
\begin{equation*}
 \left(\lambda  I + \mathcal{K}_D^*  \right) [\Delta_{\dr D} U] = 0.
\end{equation*}
Therefore, either $-\lambda \in \sigma(\mathcal{K}^*_D)$ or $\Delta_{\dr D} U=0$, which implies that $U$ is constant and $\g_{\dr D} \cdot \phi =0$.
In this case, we take the surface curl of (\ref{vecteurspropres2}) to get
\begin{equation*}
- \lambda \Delta_{\dr D}  V - \mathrm{curl}_{\dr D}  \mathcal{M}_{ D}\left[\mathrm{curl}_{ \dr D} V \right]= 0.
\end{equation*}
Using (\ref{curl}), we obtain
\begin{equation*}
\Delta_{\dr D}   \left(\lambda I- \mathcal{K}_{D}\right) [  V ]=0.
\end{equation*}
Then, 
$ \lambda V -  \mathcal{K}_{D}[V] =c$ for some constant $c$.  By replacing $V$ by 
$V^\prime = V+ \frac{c}{\lambda - 1/2}$ and using the fact that $\mathcal{K}_{D}[1]= 1/2$, we arrive at $\lambda V^\prime -  \mathcal{K}_{D}[V^\prime] = 0$. 
If $\lambda \notin \sigma(\mathcal{K}_D)$, then  $\phibf$ would be constant, which would yield a contradiction.  

Now, let $\lambda \in \sigma(\mathcal{K}_D)\setminus\{ 1/2\}$ and let $\varphi$ be an eigenvector associated with $\lambda$. From
\begin{equation*}
\left(\lambda I - \mathcal{K}_D\right) [\phi] = 0,
\end{equation*}
Taking the surface curl and using (\ref{curl}) gives
\begin{equation*}
\left(\lambda I - \mathcal{M}_D\right) [\mathrm{curl}_{\dr D}\varphi] = 0.
\end{equation*}
Either $\lambda \in \sigma (\mathcal{M}_D)$ or $\mathrm{curl}_{\dr D}\phi =0$, which means that $\phi$ is constant (\cite{buffa2002traces}). Since $\lambda \neq 1/2$, $\varphi$ cannot be constant. \end{proof}

\begin{lemm}\label{resolvantMH}
Let $\phibf \in \mathrm{H}:=\mathrm{curl}_{\dr D}\left(H^{1}(\dr D)\right)$ ($\mathrm{H}$ is the space of divergence free vectors in $L_T^2$). The following resolvent estimate holds:
\begin{equation}
\Vert \left( \lambda I - \mathcal{M}_D\right)^{-1} [\phibf] \Vert_H \leq \frac{c}{\mathrm{dist}\left(\lambda,\sigma(\mathcal{M}_D)\right)} \Vert \phibf \Vert_H.
\end{equation}

\end{lemm}
\begin{proof}
We proceed exactly as in the proof of Proposition \ref{resolvantK}. If we denote by $\langle .,.\rangle_\mathrm{H}$ the usual scalar product on $\mathrm{H}$, then we introduce a new scalar product defined by 
\begin{equation*}
\forall \phibf,\psibf \in \mathrm{H}\times \mathrm{H},\quad  \langle \phibf , \psibf\rangle_\mathcal{N} = \langle\mathcal{N}_D[\phibf] , \psibf\rangle_\mathrm{H},
\end{equation*} where $\mathcal{N}_D\big\vert_\mathrm{H}$ is the operator induced by $\mathcal{N}_D$ given in (\ref{defN}) on $\mathrm{H}$. Then, we first prove that $\mathrm{H}$ is stable by $\mathcal{N}_D$. If $\phibf \in \mathrm{H}$, then $\mathcal{N}_D[\phibf] \in  \mathrm{TH}(\mathrm{div},\dr D)$ (see \cite{colton2012inverse}) and, using the fact that for any $\mathrm{f}\in H(\mathrm{curl},\Omega)$, $\g_{\dr D} \cdot(\nubf \times \mathrm{f} ) = \nubf \cdot \g \times \mathrm{f}$, we get 
\begin{equation*}
\g_{\dr D}\cdot  \mathcal{N}_D[\phibf] = \nubf \cdot \g \times \g \mathcal{S}_D[ \g_{\dr D} \cdot(\nubf \times \phibf)] =0,
\end{equation*}  which means that $ \mathcal{N}_D[\phibf]\in \mathrm{H}$. 
For the sake of simplicity we will denote by $\mathcal{N}_D$ the induced operator on $\mathrm{H}.$
 It is easy to see that this bilinear operator is well defined, continuous and positive.
Then, $\mathcal{N}_D$ is self-adjoint \cite{colton2012inverse}. The bilinear form is positive since
\begin{equation*}
\begin{aligned}
\langle \mathcal{N}[\phibf],\phibf\rangle_\mathrm{H}&= \int_{\dr D} \mathcal{N}[\phibf](x) \cdot \phibf(x)dx ,\\
&= \int_{\dr D} \nubf(x) \times \g\mathcal{S}_D\left[\g_{\dr D} \cdot (\nubf \times\phibf)\right](x) \cdot \phibf(x) dx,\\
 &=\int_{\dr D} -\mathrm{curl}_{\dr D} \mathcal{S}_D\left[
 \mathrm{curl}_{\dr D} \phibf\right] (x) \cdot \phibf(x) dx,\\
 &= -\int_{\dr D} \mathcal{S}_D\left[\mathrm{curl}_{\dr D} \phibf\right](x) \mathrm{curl}_{\dr D} \phibf (x)  dx,\\
&=-\langle \mathcal{S}_D \left[\mathrm{curl}_{\dr D} \phibf\right], \mathrm{curl}_{\dr D} \phibf\rangle_{L^2(\dr D)}.
  \end{aligned}
\end{equation*}
If we equip $\mathrm{H}$ with this new scalar product, then we can see similarly to Proposition \ref{resolvantK}  that $\mathcal{M}_D$ is self-adjoint and therefore,  
\begin{equation*}
\forall \phibf \in \mathrm{H}, \quad \Vert \left(\lambda I-\mathcal{M}_D\right)^{-1} [\phibf] \Vert_{\mathcal{N}} \leq \frac{1}{\mathrm{dist}\left(\lambda, \sigma(\mathcal{M}_D)\right)} \Vert\phibf\Vert_\mathrm{H}.
\end{equation*}
Using the fact that $\mathcal{N}_D$ is injective and continuous on $\mathrm{H}$, we can go back to the original norm to have
\begin{equation*}
\forall \phibf \in \mathrm{H}, \quad \Vert \left(\lambda I-\mathcal{M}_D\right)^{-1}[\phibf] \Vert_\mathrm{H} \leq \frac{C}{\mathrm{dist}\left(\lambda, \sigma(\mathcal{M}_D)\right)}\Vert\phibf\Vert_\mathrm{H},
\end{equation*}
which completes the proof. \end{proof}

\begin{prop}\label{resolvantM}
Let $\lambda \in \mathbb{C}\setminus [-\frac{1}{2},\frac{1}{2}]$. There exists a positive constant $C$ such that
\begin{equation}
\forall \phibf \in L^2_T(\dr D), \quad \Vert \left( \lambda I - \mathcal{M}_D\right)^{-1} [\phibf] \Vert_{L^2_T(\dr D)} \leq \frac{C}{\mathrm{dist}(\lambda,\sigma(\mathcal{M}_D))}\Vert\phibf\Vert_{L^2_T(\dr D)}.
\end{equation}	

\end{prop}
\begin{proof}
Let $\psibf, \phibf \in \left(L^2_T(\dr D)\right)^2$ be such that 
\begin{equation}\label{eqphipsi}
\left(\lambda I - \mathcal{M}_D\right)[\psibf] = \phibf.
\end{equation}
Using Helmholtz decomposition (\ref{helmdecom}), we can write 
\begin{equation*}
\psibf = \g_{\dr D} U + \mathrm{curl}_{\dr D} V,
\end{equation*} with $U \in H^1(\dr D)$ and $V\in H^{1/2} (\dr D)$. Taking the surface divergence of (\ref{eqphipsi}), together with using (\ref{divM0}), (\ref{curl}), and the fact that $\g_{\dr D} \cdot \mathrm{curl}_{\dr D} f=0,\  \forall f$, yields
\begin{equation*}
\left(\lambda I - \mathcal{K}^*_D\right) \left[\Delta_{\dr D} U\right] = \g_{\dr D}\cdot \phibf,
\end{equation*}
which can be written as 
\begin{equation}
\Delta_{\dr D} U = \left(\lambda I - \mathcal{K}_D^*\right)^{-1} \left[\g_{\dr D} \phibf\right].
\end{equation}

\bigskip

Now we deal with the curl part.
If we apply  $\mathcal{N}_D$ on (\ref{eqphipsi})  we get by using (\ref{calderon}) together with Lemma \ref{kernelN} that
\begin{equation*}
\left(\lambda I - \mathcal{M}^*_D\right) \mathcal{N}_D 
[\mathrm{curl}_{\dr D} V] = \mathcal{N}_D[\phibf],
\end{equation*}
or equivalently, 
\begin{equation}\label{curlpart}
 \mathcal{N}_D [\mathrm{curl}_{\dr D} V] =\left(\lambda I - \mathcal{M}^*_D\right)^{-1} \mathcal{N}_D [\phibf].
\end{equation}
From the Helmholtz decomposition of $\phibf$: $\phibf = \g_{\dr D} \phibf_1 + \mathrm{curl}_{\dr D} \phibf_2$, (\ref{curlpart}) becomes
\begin{equation}\label{curlpart2}
\mathcal{N}_D[\mathrm{curl}_{\dr D} V] =\left(\lambda I - \mathcal{M}^*_D\right)^{-1} \mathcal{N}_D\left[\mathrm{curl}_{\dr D} \phibf_2\right].
\end{equation}
Now, we can work in the function space $\mathrm{H}=\mathrm{curl}_{\dr D} H^{1/2}(\dr D)$. We denote by $\widetilde{\mathcal{N}}_D$ the operator induced by $\mathcal{N}_D$ on $\mathrm{H}$ and by $R(\widetilde{\mathcal{N}}_D) \subset \mathrm{H}$ the range of the induced operator. $\mathcal{M}_D$ also induces an operator $\widetilde{\mathcal{M}}_D$ on $\mathrm{H}$; see the proof of (\ref{calderon}).

Next, we want to make sure that $\left(\lambda I - \widetilde{\mathcal{M}}^*_D\right)^{-1} \widetilde{\mathcal{N}}_D [\mathrm{curl}_{\dr D} V ] $ belongs to $R(\widetilde{\mathcal{N}}_D)$ so that we can apply $\widetilde{\mathcal{N}}_D$'s left inverse (recall from Lemma \ref{kernelN} that $\widetilde{\mathcal{N}}_D$ is injective).
For doing so, we show that the range of $\widetilde{\mathcal{N}}_D$ is stable by $\left(\lambda I - \widetilde{\mathcal{M}}^*_D\right)^{-1}$.
Take ${f}=\widetilde{\mathcal{N}}_D[{g}] \in R(\widetilde{\mathcal{N}}_D)$. Then,
\begin{equation*}
\begin{aligned}
\left(\lambda I - \widetilde{\mathcal{M}}^*_D\right)^{-1}[{f}] \in R(\widetilde{\mathcal{N}}_D) &\Leftrightarrow \exists {h}\in \mathrm{H}, \quad  \left(\lambda I - \widetilde{\mathcal{M}}^*_D\right)^{-1} \widetilde{\mathcal{N}}_D [{g}]=\widetilde{\mathcal{N}}_D [{h}]\\
 &\Leftrightarrow \exists {h}\in \mathrm{H}, \quad \widetilde{\mathcal{N}}_D [{g}] = \left(\lambda I - \widetilde{\mathcal{M}}_D^*\right)\widetilde{\mathcal{N}}_D [{h}]\\
&\Leftrightarrow  \exists {h}\in \mathrm{H}, \quad  \widetilde{\mathcal{N}}_D [{g}]= \widetilde{\mathcal{N}}_D \left(\lambda I - \widetilde{\mathcal{M}}_D\right) [{h}]
\\ &\Leftrightarrow  \exists {h}\in \mathrm{H}, \quad   {g}=  \left(\lambda I - \widetilde{\mathcal{M}}_D\right)[{h}] \quad (\mbox{ by injectivity of } \widetilde{\mathcal{N}}_D)
\\ &\Leftrightarrow  \exists {h}\in \mathrm{H}, \quad   \left(\lambda I - \widetilde{\mathcal{M}}_D\right)^{-1} [{g}]=   {h}. 
\end{aligned}
\end{equation*}
So we have the stability of $R(\widetilde{\mathcal{N}}_D)$ by $\widetilde{\mathcal{M}}_D$ and
\begin{equation}\label{NresolvantM}
\widetilde{\mathcal{N}}_D^{-1}\left(\lambda I - \widetilde{\mathcal{M}}^*_D\right)^{-1} \widetilde{\mathcal{N}}_D = \left(\lambda I - \widetilde{\mathcal{M}}_D\right)^{-1}.
\end{equation}
Applying this to (\ref{curlpart2}) we get
\begin{equation*}
\mathrm{curl}_{\dr D} V = \left(\lambda I - \widetilde{\mathcal{M}}_D\right)^{-1} [\mathrm{curl}_{\dr D} \phibf_2].
\end{equation*}
Using Lemma \ref{resolvantMH} we get the desired result. \end{proof}

An important remark is on order. In view of Proposition \ref{spectrumequality}, we may think that both $\sigma(\mathcal{K}^*_D)$ and $- \sigma(\mathcal{K}^*_D)$ contribute to the plasmonic resonances at the quasi-static regime. However, as it will be seen in the next section, only $\sigma(\mathcal{K}^*_D)$ contributes to the resonances at the zero size limit.  $- \sigma(\mathcal{K}^*_D)$ contributes to higher-order resonances in terms of the particle size. Moreover, it is worth stating the following estimate which follows immediately from the proof of Proposition \ref{spectrumequality}: 
\begin{equation}
\label{important}
\Vert \left( \lambda I - \mathcal{M}_D\right)^{-1} [\mathrm{curl}_{\partial D} \phi] \Vert_{L^2(\dr D)} \leq \frac{C}{\mathrm{dist}(\lambda,\sigma(\mathcal{K}^*_D))}\Vert\phi\Vert_{H^1(\dr D)}
\end{equation}
for some constant $C$.

\section{Small  volume expansion} \label{sec22}

The aim of this section is to prove Theorems \ref{theo2f} and \ref{theo3f}. 

\subsection{Layer potential formulation}
For a given plane wave solution $(\bE^i, \bH^i)$ to the Maxwell
equations
\begin{equation*}
 \left \{
 \begin{array}{ll}
\nabla\times{\bE^i} = i\omega\mu_m{\bH^i} \quad &\mbox{in } \mathbb{R}^3,\\
\nabla\times{\bH^i} = -i\omega\varepsilon_m {\bE^i}\quad &\mbox{in
} \mathbb{R}^3,
 \end{array}
 \right .
 \end{equation*}
let $(\bE,\bH)$ be the solution to the following Maxwell equations: 
\begin{equation}
\label{eq:maxwell} \left\{
\begin{array}{ll}
 \nabla \times \bE = i \omega \mu \bH &  \mbox{in} \quad \mathbb{R}^3\setminus \dr D, \\
 \nabla  \times \bH= - i \omega \varepsilon \bE & \mbox{in} \quad \R^3\setminus \dr D, \\
 {[}{\nu} \times \bE]= [{ \nu} \times \bH] = 0 & \mbox{on} \quad \dr D,
\end{array}
\right.
\end{equation}  subject to the Silver-M\"{u}ller radiation condition:
 $$\lim_{|x|\rightarrow\infty} |x| (\sqrt{\mu} (\bH- \bH^i) \times\hat{x}-\sqrt{\varepsilon} (\bE-\bE^i))=0,$$
 where $\hat{x} = x/|x|$. Here, $[{\nubf} \times{\bE}]$ and $[{\nubf} \times{\bH}]$ denote the jump of ${\nubf} \times{\bE}$
and ${\nubf} \times{\bH}$ along $\dr D$, namely,
 \begin{equation*}
 [\nubf\times{\bE}]=(\nubf\times \bE)^+ -(\nubf \times \bE)^-,\quad [\nubf\times{\bH}]=(\nubf\times \bH)^+ -(\nubf\times \bH)^-.
  \end{equation*} 

Using the layer potentials defined in section \ref{sec2}, the solution to  (\ref{eq:maxwell}) can be represented as
 \begin{equation}
\label{represent} \ds\bE (x)= \left \{
 \begin{array}{ll}
\ds \bE^i(x) + \mu_m \nabla \times \mathcal{A}_D^{k_m}
[\phibf](x) +
\nabla\times\nabla\times\mathcal{A}_D^{k_m} [\psibf](x) ,\quad &x \in \mathbb{R}^3 \setminus \overline{D},\\
\nm \ds\mu_c \nabla \times \mathcal{A}_D^{k_c} [\phibf](x) +
\nabla\times\nabla\times\mathcal{A}_D^{k_c} [\psibf](x) ,\quad &x
\in D,
 \end{array}
 \right .
 \end{equation}
and
\begin{equation}
\bH(x) = -\frac{i}{\omega \mu}\bigr(\nabla \times \bE\bigr)(x),\quad x \in \mathbb{R}^3\setminus \dr D,
\end{equation}
where the pair $(\phibf, \psibf) \in TH({\rm div}, \dr D)
\times TH({\rm div}, \dr D)$ is the unique solution to \begin{equation}
\label{phi_psi}
\begin{bmatrix}
\ds\frac{\mu_c+\mu_m}{2}I + \mu_c \mathcal{M}_D^{k_c} -\mu_m
\mathcal{M}_D^{k_m} &
\ds\mathcal{L}_D^{k_c} - \mathcal{L}_D^{k_m} \\
\ds\mathcal{L}_D^{k_c} - \mathcal{L}_D^{k_m} & \ds \left(
\frac{k_c^2}{2 \mu_c} + \frac{k_m^2}{2 \mu_m}\right)I +
\frac{k_c^2}{\mu_c}\mathcal{M}_D^{k_c} -
\frac{k_m^2}{\mu_m}\mathcal{M}_D^{k_m}
\end{bmatrix}
\begin{bmatrix}
\phibf \\ \psibf
\end{bmatrix}
= \left.\begin{bmatrix}
 {\nu}\times \bE^i\\
i \omega {\nu} \times \bH^i
\end{bmatrix}\right|_{\dr D}.
\end{equation} The invertibility of the system of equations (\ref{phi_psi})
on $TH({\rm div}, \dr D) \times TH({\rm div}, \dr D)$ was proved in
\cite{T}. Moreover, there exists a constant $C=C(\varepsilon, \mu,
\omega)$ such that 
\begin{equation}
\label{phi_psi_Ein_Hin} \ds\| \phibf
\|_{TH({\rm div},\dr  D)}+ \| \psibf \|_{TH({\rm div},\dr  D)} \leq
C\bigr(\| \bE^i \times {\nu} \|_{TH({\rm div},\dr  D)}+ \|\bH^i
\times {\nu} \|_{TH({\rm div},\dr  D)}\bigr). 
\end{equation}
%Denote by $\lambda_{\varepsilon}$ and $\lambda_{\mu}$ the
%electric permittivity and magnetic permeability contrasts: \begin{equation}
%\lambda_{\varepsilon}=\frac{\varepsilon_c +\varepsilon_m}{2(\varepsilon_c-\varepsilon_m)}
%\quad \mbox{and} \quad
%\lambda_{\mu}=\frac{\mu_c+\mu_m}{2(\mu_c-\mu_m)}
%\end{equation}
%which are important parameters for later analysis.

\subsection{Derivation of the asymptotic formula}
We will need the following notation. For a multi-index $\alpha \in
\mathbb{N}^3$, let $x ^\alpha = x_1^{\alpha_1} x_2^{\alpha_2}
x_3^{\alpha_3}$, $\partial^\alpha = \partial_1^{\alpha_1} \partial_2^{\alpha_2}
\partial_3^{\alpha_3},$ with $\partial_j = \partial /
\partial x_j$.

Let $D=\delta B+ z$, where $B$ is a $\mathcal{C}^{1,\eta}$ ($0<\eta<1$)
domain containing the origin. For any $y\in \dr D$, let $\widetilde{y}=\frac{y-z}{\delta} \in \dr  B$. Denote by
$\widetilde{\phibf}(\widetilde{y})= \phibf(y)$ and $\widetilde{\psibf}(\widetilde{y})= \psibf(y)$.

\subsubsection{Asymptotics for the operators}
We have the following expansions for $ \mathcal{M}_D^k$ and $\mathcal{L}_D^k$.
\begin{prop}\label{asymptotiqueM}
Let $\phibf \in L_T^2(\dr D)$. As $\delta \rightarrow 0$, we have
\begin{equation}
\mathcal{M}_D^k[\phibf](x)=\mathcal{M}_B[\widetilde{\phibf}](\widetilde{x}) + O(\delta^2).
\end{equation}

\end{prop}
\begin{proof}
Let $x\in \dr D$, and write $\widetilde{x}=\frac{x-z}{\delta}$. We have
\begin{equation*}
\mathcal{M}_D^k[\phibf](\delta\widetilde{x}+z)=-\frac{1}{4\pi\delta}\int_{\dr D} \nubf_D(\delta\widetilde{x}+z) \times \left( \nabla_{\widetilde{x}} \times \left( \frac{e^{i k \vert \delta \widetilde{x}+ z - y \vert }}{\vert \delta \widetilde{x}+ z -y \vert } \phibf(y) \right) \right)d\sigma(y).
\end{equation*}
Changing $y$ by $\widetilde{y} = \frac{y-z}{\delta}$ in the integral we get
\begin{equation*}
\mathcal{M}_D^k[\phibf](\delta\widetilde{x}+z)=- \frac{1}{4 \pi \delta}\int_{\dr B} \nubf_D(\delta\widetilde{x}+z) \times \left( \nabla_{\widetilde{x}} \times \left( \frac{e^{i k \delta \vert  \widetilde{x}-\widetilde{y} \vert }}{\delta \vert  \widetilde{x}-\widetilde{y} \vert  } \widetilde{\phibf}(\widetilde{y}) \right) \right) \delta^2 d\sigma(\widetilde{y}).
\end{equation*}
Since $ \forall x \in \dr D, \nubf_D(x)=\nubf_B(\frac{x-z}{\delta})$, 
\begin{equation*}
\mathcal{M}_D^k[\phibf](x)=\mathcal{M}_B^{\delta k}[\widetilde{\phibf}](\widetilde{x}).
\end{equation*}
For any $\widetilde{x}\in \delta B$, it follows that
\begin{equation*}
\mathcal{M}_B^{\delta k}[\widetilde{\phibf}](\widetilde{x})=\mathcal{M}_B[\widetilde{\phibf}](\widetilde{x}) +\int_{\dr B}\nubf_B (\widetilde{x}) \times \left( \g_{\widetilde{x}} \times \left( i k\delta  \right)\right)  +O \left(\delta^2\right),
\end{equation*} which gives the result.\end{proof}

\begin{prop}\label{asymptotiqueL} Let $\phibf \in TH(\rm div, \dr D)$.
For any $y\in \dr D$, we have
\begin{multline}
\mathcal{L}_D^{k_m}[\phibf](y)-\mathcal{L}_D^{k_c}[\phibf](y) = \\ \delta(k_m^2-k_c^2)\nubf_B(\widetilde{y})\times\left(\mathcal{A}_B[\widetilde{\phi}](\widetilde{y}) + \frac{1}{8\pi}\int_{\dr B} \frac{\widetilde{y}-\widetilde{y}'}{\vert \widetilde{y}- \widetilde{y}'\vert } \left(\g_{\dr B}\cdot \widetilde{\phibf}(\widetilde{y}')\right) d\sigma(\widetilde{y}')\right) + O\left(\delta^2)\right).
\end{multline}
\end{prop}
\begin{proof}
Note that, for $y\in \dr D$,
\begin{equation*}
\mathcal{A}_D^k[\phi](y)=\delta \mathcal{A}_B^{\delta k}[\widetilde{\phi}](\widetilde{y})
\end{equation*}
and 
\begin{equation*}
\g_{\dr D}\mathcal{S}_D^{ k}[\g_{\dr B}\cdot \phibf](y)=\frac{1}{\delta} \g_{\dr B}\mathcal{S}_B^{\delta k}[\g_{\dr B}\cdot \widetilde{\phibf}](\widetilde{y}) .
\end{equation*}
We can expand 
\begin{equation*}
\mathcal{A}_B^{\delta k}[\widetilde{\phibf}](\widetilde{y}) =  \mathcal{A}_B[\widetilde{\phibf}](\widetilde{y}) + O(\delta). 
\end{equation*}
We also have
\begin{multline*}
\g_{\dr B}\mathcal{S}_B^{\delta k}[\g_{\dr B}\cdot \widetilde{\phibf}](\widetilde{y}) = - \frac{1}{4 \pi} \\ \times \g_{\dr B} \int_{\dr B} \frac{1}{\vert \widetilde{y}- \widetilde{y}'\vert }\left( 1+ k\delta \vert \widetilde{y}- \widetilde{y}'\vert  -\frac{1}{2} k^2\delta^2\vert \widetilde{y}- \widetilde{y}'\vert^2   +O(\delta^3 \vert \widetilde{y}- \widetilde{y}'\vert^3)\right) \g_{\dr B}\cdot \widetilde{\phibf}(\widetilde{y}') d\sigma(\widetilde{y}')
\end{multline*}
and
\begin{multline*}
\g_{\dr B}\mathcal{S}_B^{\delta k}[\g_{\dr B}\cdot \widetilde{\phibf}](\widetilde{y}) = - \frac{1}{4 \pi} \g_{\dr B} \int_{\dr B}\frac{1}{\vert \widetilde{y}- \widetilde{y}'\vert } \g_{\dr B}\cdot \widetilde{\phibf}(\widetilde{y}') d\sigma(\widetilde{y}')\\ - \frac{1}{2}\g_{\dr B} \int_{\dr B}\vert \widetilde{y}- \widetilde{y}'\vert  \g_{\dr B}\cdot \widetilde{\phibf}(\widetilde{y}') + O(\delta^3).
\end{multline*}
Now, since $\displaystyle{\forall f\in L^2(\dr B),\ \mathcal{S}_B[f]\big|_{\ol{B}}\in \mathcal{C}^1(\ol{B}), \ \mathcal{S}_B[f]\big|_{\mathbb{R}^3\setminus B }\in \mathcal{C}^1(\mathbb{R}^3\setminus B)}$ and the tangential component of the gradient of $\mathcal{S}[f]$ is continuous across $\partial B$, we can state that $$\forall \widetilde{y}\in \dr B,  \ \nubf_B(\widetilde{y}) \times \g_{\dr B}\mathcal{S}_B [f]\big|_{\dr B} (\widetilde{y})= \nubf_ D(\widetilde{y}) \times \g \mathcal{S}_B[f]\big|_{\mathbb{R}^3\setminus B} (\widetilde{y})=\nubf_ D(\widetilde{y}) \times \g \mathcal{S}_B[f]\big|_{\ol{B}} (\widetilde{y}).$$
Then we can write 
\begin{multline*}
\forall \widetilde{y}\in \dr B,  \ \nubf_B(\widetilde{y}) \times\g_{\dr B}\mathcal{S}_B^{\delta k}[\g_{\dr B}\cdot \widetilde{\phibf}](\widetilde{y}) = - \frac{1}{4 \pi} \nubf_B(\widetilde{y}) \times \bigg[  \g_{\dr B} \int_{\dr B}\frac{1}{\vert \widetilde{y}- \widetilde{y}'\vert } \g_{\dr B}\cdot \widetilde{\phibf}(\widetilde{y}') d\sigma(\widetilde{y}') \\-\frac{1}{2} \int_{\dr B} \frac{\widetilde{y}- \widetilde{y}'}{\vert \widetilde{y}- \widetilde{y}' \vert }\g_{\dr B}\cdot \widetilde{\phibf}(\widetilde{y}') d\sigma(\widetilde{y}) + O(k^3\delta^3)\bigg].
\end{multline*}
The proof is then complete. \end{proof} 

\subsubsection{Far-field expansion}
Define $\widetilde{\phibf}_\beta$ and $\widetilde{\psibf}_\beta$ for every $\beta \in \mathbb{N}^3$ by
\begin{equation}
\mathcal{W}_B^{\delta} \left[
\begin{array}{c}
 \widetilde{\phibf}_{\beta} \\
 \widetilde{\psibf}_{\beta}
\end{array}
\right]= \left[
\begin{array}{c}
 { \nu}(\widetilde{y}) \times (\widetilde{y}^{\beta}\dr^{\beta} {E}^i(z))  \\
 i\omega { \nu}(\widetilde{y})\times (\widetilde{y}^{\beta}\dr^{\beta} {H}^i(z))
\end{array}
\right]\end{equation}
with \begin{equation}
\mathcal{W}_B^{\delta}=\left[
\begin{array}{cc}
\frac{\mu_m+\mu_c}{2}I+ \mu_c\mathcal{M}_B^{\delta k_c} - \mu_m\mathcal{M}_B^{\delta k_m}
 & \mathcal{L}_{B,\delta}^{k_c} - \mathcal{L}_{B,\delta}^{k_m}\\
\mathcal{L}_{B,\delta}^{k_c} - \mathcal{L}_{B,\delta}^{k_m}&
\omega^2(\frac{\varepsilon_m+\varepsilon_c}{2}I+
\varepsilon_c\mathcal{M}_B^{\delta k_c} - \varepsilon_m\mathcal{M}_B^{\delta k_m})
\end{array}
\right].
\end{equation}

Using (\ref{represent}) we have the following expansion for $\bE(x)$ for $x$ far away from $z$:
\begin{multline}
\bE(x)= \bE^i(x) + \sum_{\vert \alpha \vert =0}^\infty \sum_{\vert \beta\vert =0}^\infty \delta^{2+\vert \alpha \vert +\vert \beta \vert } \frac{(-1)^{\vert \alpha\vert}}{\alpha ! \beta !} \bigg( \mu_m \g\dr^\alpha\Gamma^{k_m} (x-z)\times \int_{\dr B} \widetilde{y}^\alpha \widetilde{\phibf}_\beta(\widetilde{y}) d\sigma(\widetilde{y})\\
+ \g \times \g \dr^\alpha \Gamma^{k_m}(x-z) \times  \int_{\dr B} \widetilde{y}^\alpha \widetilde{\psibf}_\beta(\widetilde{y}) d\sigma(\widetilde{y})\bigg).
\end{multline}

For $\beta \in \mathbb{N}^3$, define the tensors by
\begin{equation}\label{eq:tns}
\mathrm{M}^e_{\alpha,\beta}:=\int_{\dr B}\widetilde{y}^\alpha\widetilde{\psibf}_{\beta}d\sigma(\widetilde{y})
\quad \mbox{and} \quad \mathrm{M}^h_{\alpha,\beta}:=\int_{\dr
B}\widetilde{y}^{\alpha}\widetilde{\phibf}_{\beta}d\sigma(\widetilde{y}).
\end{equation}

The following lemma holds.
\begin{lemm}\label{le:exp1} For $x\in\mathbb{R}^3\setminus \overline{D}$, we have
\begin{multline}
\bE(x) = \bE^i(x) + \sum_{|\alpha|=0}^{\infty}\sum_{|\beta|=0}^{\infty}
\delta^{2+|\alpha|+|\beta|}\frac{(-1)^{|\alpha|}}{\alpha !\beta!}
\Big(\mu_m\nabla\dr^{\alpha}\Gamma^{k_m}(x-z)\times \mathrm{M}^h_{\alpha,\beta} \\
 +\nabla\times\nabla\dr^{\alpha}\Gamma^{k_m}(x-z)\times
\mathrm{M}^e_{\alpha,\beta}\Big).
\end{multline}

\end{lemm}

\subsubsection{Asymptotics for the potentials}
\begin{prop}\label{estimatephi_beta_n}
Let $\beta\in \mathbb{N}^3$. We can write the following expansions for $\widetilde{\phibf}_\beta$ and $\widetilde{\psibf}_\beta:$
 $$
 \widetilde{\phibf}_\beta= \sum_{n=0}^{\infty} \delta^n \widetilde{\phibf}_{\beta,n}, \qquad 
 \widetilde{\psibf}_\beta= \sum_{n=0}^{\infty} \delta^n \widetilde{\psibf}_{\beta,n}.
 $$ Moreover, there exists a $C\geq 0$ depending on $B$, $\beta$, ${E}$, and ${H}$ such that
\begin{equation}
\begin{aligned}
\forall n\in \mathbb{N}, \Vert \widetilde{\phibf}_{\beta,n}\Vert_{\mathrm{TH}(\mathrm{div},\dr B)}\leq& C^{(n+1)}\left(\frac{1}{ \mathrm{dist}(\lambda_\mu, \sigma(\mathcal{M}_B))}\right)^{\lfloor n/2\rfloor } \left(\frac{1}{ \mathrm{dist}(\lambda_\varepsilon,\sigma(\mathcal{M}_B))}\right)^{\lfloor n/2+1\rfloor} ,\\
\forall n\in \mathbb{N}, \Vert \widetilde{\psibf}_{\beta,n}\Vert_{\mathrm{TH}(\mathrm{div},\dr B)}\leq& C^{(n+1)}\left(\frac{1}{ \mathrm{dist}(\lambda_\varepsilon, \sigma(\mathcal{M}_B))}\right)^{\lfloor n/2\rfloor } \left(\frac{1}{ \mathrm{dist}(\lambda_\mu,  \sigma(\mathcal{M}_B))}\right)^{\lfloor n/2+1\rfloor}  .
\end{aligned}
\end{equation}
\end{prop}
\begin{proof} We proceed  by induction. Using Propositions \ref{asymptotiqueM} and \ref{asymptotiqueL} we find that 
\begin{equation}\label{eqphipsi00}
\begin{aligned}
\widetilde{\phibf}_{\beta,0}&= (\mu_c-\mu_m)^{-1}(\lambda_\mu I - \mathcal{M}_B)^{-1}\left[ \nubf(\widetilde{y}) \times (\widetilde{y}^\beta \dr^\beta {E}(z)\right],\\
\widetilde{\psibf}_{\beta,0}&=i\omega^{-1} (\varepsilon_c-\varepsilon_m)^{-1}(\lambda_\varepsilon I - \mathcal{M}_B)^{-1}\left[ \nubf(\widetilde{y}) \times (\widetilde{y}^\beta \dr^\beta {H}(z)\right].
\end{aligned}
\end{equation}
Note that $\g_{\dr B} \cdot \widetilde{\phibf}_{\beta,0}=0$ for $\beta=0$.
Indeed, 
\begin{equation*}
\g_{\dr B}\cdot \widetilde{\phibf} = (\mu_c-\mu_m)^{-1}(\lambda_\mu I - \mathcal{K}^*_B)^{-1}\left[\g_{\dr B}\cdot \left( \nubf(\widetilde{y}) \times (\widetilde{y}^\beta \dr^\beta {E}(z)\right)\right],
\end{equation*}
and
\begin{equation*}\begin{aligned}
\g_{\dr B} \cdot  \left( \nubf(\widetilde{y}\right) \times \left(\widetilde{y}^\beta \dr^\beta {E}(z)\right) =& \nubf(\widetilde{y}) \cdot \left( \g \times [\widetilde{y}^\beta {E}(z) ]\right)\\
=& 0.
\end{aligned}
\end{equation*}
In the same way we have  $\g_{\dr B} \cdot \widetilde{\psibf}_{\beta,0}=0$  for $\beta=0$.
Using Proposition \ref{resolvantM}, we get the result.

For the first-orders the equations satisfied by $\widetilde{\phibf}_{\beta,1}$ and $\widetilde{\psibf}_{\beta,1}$ are 
\begin{equation}\label{eqphi1psi1}
\begin{aligned}
(\mu_c-\mu_m)(\lambda_\mu I - \mathcal{M}_B)[\widetilde{\phibf}_{\beta,1}]+ (k_c^2-k_m^2)\nubf(\widetilde{y}) \times \mathcal{A}_B[\widetilde{\psibf}_{\beta,0}]&= 0,\\
\nm
\omega^2(\varepsilon_c-\varepsilon_m)(\lambda_\varepsilon I - \mathcal{M}_B)[\widetilde{\psibf}_{\beta,1}]+ (k_c^2-k_m^2)\nubf(\widetilde{y}) \times \mathcal{A}_B[\widetilde{\phibf}_{\beta,0}]&= 0.
\end{aligned}
\end{equation}
The fact that $\mathcal{A}_B$ is bounded together with Proposition \ref{resolvantM} gives the estimate of $\Vert \widetilde{\phibf}_{\beta,1}\Vert_{L^2_T(\dr B)}$ and $\Vert\widetilde{\phibf}_{\beta,1}\Vert_{L^2_T(\dr B)}$. If we take the surface divergence of  (\ref{eqphi1psi1}), we get
\begin{equation*}
\begin{aligned}
(\mu_c-\mu_m)(\lambda_\mu I - \mathcal{K}^*_B)[\g_{\dr B} \cdot \widetilde{\phibf}_{\beta,1}]+ (k_c^2-k_m^2)\g_{\dr B}\cdot \left(\nubf(\widetilde{y}) \times \mathcal{A}_B[\widetilde{\psibf}_{\beta,0}]\right)&= 0,\\
\omega^2(\varepsilon_c-\varepsilon_m)(\lambda_\varepsilon I - \mathcal{K}^*_B)[\g_{\dr B}\cdot \widetilde{\psibf}_{\beta,1}]+ (k_c^2-k_m^2)\g_{\dr B}\cdot \left(\nubf(\widetilde{y}) \times \mathcal{A}_B[\widetilde{\phibf}_{\beta,0}]\right)&= 0.
\end{aligned}
\end{equation*}
Since $\g_{\dr B}\cdot \left(\nubf(\widetilde{y}) \times \mathcal{A}_B[\widetilde{\psibf}_{\beta,0}]\right) =\nubf(\widetilde{y})\cdot\left( \g \times \mathcal{A}_B[\widetilde{\phibf}_{\beta,0}]\right) $ and ${f} \mapsto \nubf \cdot \g \times \mathcal{A}_B[{f}]$ is bounded from $L^2_T(\dr B)$ into $L^2(\dr B)$ , we can estimate the $L^2$ norm of $\g_{\dr B}\cdot  \widetilde{\phibf}_{\beta,1}$ as follows
\begin{equation*}
\left\Vert \frac{1}{\mu_c-\mu_m} \left( \lambda_\mu - \mathcal{K}_B^*\right)^{-1}\left[\nubf(\widetilde{y})\cdot\left( \g \times \mathcal{A}_B[\widetilde{\phibf}_{\beta,0}]\right)\right]\right\Vert_{L^2(\dr B)} \leq  \frac{c}{ \mathrm{dist}(\lambda_\mu, \sigma(\mathcal{K}_B))} \Vert \widetilde{\phibf}_{\beta,0}\Vert_{L^2_T}.
\end{equation*}
From Proposition \ref{spectrumequality} we get the result. The estimate for $\Vert \g_{\dr B} \cdot \widetilde{\psibf}_{\beta,1}\Vert_{L^2}$ is obtained in the same way.

Now, fix $n \in \mathbb{N}^*$ ; $\widetilde{\phibf}_{\beta,n+1}$ and $\widetilde{\psibf}_{\beta,n+1}$ satisfy the following system:
\begin{equation*}
\begin{aligned}
(\mu_c-\mu_m)(\lambda_\mu I - \mathcal{M}_B)[\widetilde{\phibf}_{\beta,i+1}]+ (k_c^2-k_m^2)\nubf(\widetilde{y}) \times\left( \mathcal{A}_B[\widetilde{\psibf}_{\beta,i}] + \mathcal{B}_B [\widetilde{\psibf}_{\beta,i}]\right) &= 0,\\
\omega^2(\varepsilon_c-\varepsilon_m)(\lambda_\varepsilon I - \mathcal{M}_B)[\widetilde{\psibf}_{\beta,0}]+ (k_c^2-k_m^2)\nubf(\widetilde{y}) \times\left( \mathcal{A}_B[\widetilde{\phibf}_{\beta,i}] + \mathcal{B}_B [\widetilde{\phibf}_{\beta,i}]\right)&= 0,
\end{aligned}
\end{equation*}
where the operator $\mathcal{B}_B$ is defined by
\begin{equation*}
\begin{aligned}
\mathrm{TH}(\mathrm{div},\dr B)\longrightarrow & \mathrm{TH}(\mathrm{div},\dr B)\\
{f}\longmapsto & \frac{1}{8\pi}\int_{\dr B} \frac{\widetilde{y}-\widetilde{y}'}{\vert \widetilde{y}- \widetilde{y}'\vert } \left(\g_{\dr B}\cdot {f}(\widetilde{y}')\right) d\sigma(\widetilde{y}').
\end{aligned}
\end{equation*}
The operator $\mathcal{B}_B$ is bounded, and we can get the norm estimates for  $\widetilde{\phibf}_{\beta,n+1}$ , $\widetilde{\psibf}_{\beta,n+1}$,  $\g_{\dr B} \cdot \widetilde{\phibf}_{\beta,n+1}$ and $\g_{\dr B} \cdot \widetilde{\psibf}_{\beta,n+1}$, as before. \end{proof}

It is worth noticing that the following estimate follows immediately from 
(\ref{important}). We have
\begin{equation}
\label{important2}
\begin{aligned}
 \Vert \widetilde{\phibf}_{\beta,0}\Vert_{\mathrm{TH}(\mathrm{div},\dr B)}\leq& C \frac{1}{ \mathrm{dist}(\lambda_\varepsilon,\sigma(\mathcal{K}^*_B))},\\
\Vert \widetilde{\psibf}_{\beta,0}\Vert_{\mathrm{TH}(\mathrm{div},\dr B)}\leq& C \frac{1}{ \mathrm{dist}(\lambda_\mu,  \sigma(\mathcal{K}^*_B))}.
\end{aligned}
\end{equation}

\subsubsection{Derivation of the leading-order potentials}

By Lemma \ref{le:exp1}, for $x\in \mathbb{R}^3\setminus\overline{D}$,
\begin{equation}\label{eq:expansion01}
\begin{multlined}
\bE(x)= \bE^i(x)+ \delta^2\Big(\mu_m\nabla\Gamma^{k_m}(x-z)\times \mathrm{M}^h_{0,0} +\nabla\times\nabla\Gamma^{k_m}(x-z)\times
\mathrm{M}^e_{0,0}\Big)\\
\quad + \delta^3\Big(\mu_m\nabla\Gamma^{k_m}(x-z)\times \sum_{j=1}^3\mathrm{M}^h_{0,j} +\nabla\times\nabla\Gamma^{k_m}(x-z)\times
\sum_{j=1}^3\mathrm{M}^e_{0,j}\Big)\\
\quad - \delta^3\Big(\mu_m\sum_{j=1}^3\nabla\dr_j\Gamma^{k_m}(x-z)\times \mathrm{M}^h_{j,0} +\nabla\times\sum_{j=1}^3\nabla\dr_j\Gamma^{k_m}(x-z)\times
\mathrm{M}^e_{j,0}\Big)\\+O(\delta^4).
\end{multlined}
\end{equation}
We start by computing $\mathrm{M}^h_{0,0}$:
\begin{equation*}
\begin{array}{lll}
\ds \mathrm{M}^h_{0,0}&=&\ds \int_{\dr B} \widetilde{\phibf}_0(\widetilde{y}) d\sigma(\widetilde{y}),\\
\nm 
&=&\ds \int_{\dr B}\widetilde{\phibf}_0(\widetilde{y})  \g \widetilde{y} d\sigma(\widetilde{y}) ,\\
\nm
&=&\ds \int_{\dr B}\widetilde{y}\g_{\dr B}\cdot \widetilde{\phibf}_0(\widetilde{y}) d \sigma(\widetilde{y}).
\end{array}
\end{equation*}
Now, using the expansion of $\widetilde{\phibf}$ given in Proposition \ref{estimatephi_beta_n} we have
\begin{equation*}
\mathrm{M}^h_{0,0}=\int_{\dr B}\widetilde{y}\g_{\dr B}\cdot \widetilde{\phibf}_{0,0}(\widetilde{y}) d \sigma(\widetilde{y}) + \int_{\dr B}\widetilde{y}\g_{\dr B}\cdot \widetilde{\phibf}_{0,1}(\widetilde{y}) d \sigma(\widetilde{y}) + O(\delta^2).
\end{equation*}
Recall (\ref{eqphipsi00}) for $\beta=0$:
\begin{equation*}
\begin{aligned}
\widetilde{\phibf}_{0,0}&= (\mu_c-\mu_m)^{-1}(\lambda_\mu I - \mathcal{M}_B)^{-1}\left[ \nubf(\widetilde{y}) \times  {E}^i(z)\right]\\
\widetilde{\psibf}_{0,0}&=i\omega^{-1} (\varepsilon_c-\varepsilon_m)^{-1}(\lambda_\varepsilon I - \mathcal{M}_B)^{-1}\left[ \nubf(\widetilde{y}) \times {H}^i(z)\right].
\end{aligned}
\end{equation*}
We can see, using (\ref{divM0}) and the fact that
\begin{equation*}
\g_{\dr B} \cdot \left(\bE^i(z) \times \nubf(\widetilde{y})\right)=\g_{\dr B} \cdot \left(\bH^i(z) \times \nubf(\widetilde{y})\right)=0,
\end{equation*}
that \begin{equation*}
\g_{\dr B} \cdot\widetilde{\phibf}_{0,0}=\g_{\dr B} \cdot\widetilde{\psibf}_{0,0}=0.
\end{equation*}
Now, taking the surface divergence of (\ref{eqphi1psi1}) for $\beta=0$, it follows that
\begin{equation}\label{eq:divphi1psi1}
\begin{aligned}
(\mu_c-\mu_m)(\lambda_\mu I - \mathcal{K}^*_B)[\g_{\dr B}\cdot\widetilde{\phibf}_{0,1}]+ (k_c^2-k_m^2)\g_{\dr B}\cdot \bigg(\nubf(\widetilde{y}) \times \mathcal{A}_B[\widetilde{\psibf}_{0,0}]\bigg)&= 0,\\
\omega^2(\varepsilon_c-\varepsilon_m)(\lambda_\varepsilon I - \mathcal{K}^*_B)[\g_{\dr B}\cdot\widetilde{\psibf}_{0,1}]+ (k_c^2-k_m^2)\g_{\dr B}\cdot \bigg(\nubf(\widetilde{y}) \times \mathcal{A}_B[\widetilde{\phibf}_{0,0}]\bigg)&= 0.
\end{aligned}
\end{equation}
Since $\g_{\dr B}\cdot (\nubf \times •) = \nubf \cdot (\g \times •)$ we need to study the quantities
\begin{equation*}
\nubf\cdot \g \times \mathcal{A}_B[\widetilde{\phibf}_{0,0}]
\end{equation*}
and
\begin{equation*}
\nubf\cdot \g \times \mathcal{A}_B[\widetilde{\psibf}_{0,0}].
\end{equation*}
The following lemma holds.
\begin{lemm}\label{le:phipsi} We have
\begin{equation}\label{eq:lephi}
\nabla\times\mathcal{A}_B[\widetilde{{\phi}}_{0,0}]=\left\{
\begin{array}{ll}
 \ds \frac{1}{\mu_c-\mu_m}\nabla\mathcal{S}_B\left(\lambda_\mu I - \mathcal{K}_B^*\right)^{-1}[{\nu}\cdot \bE^i(z)] & \mbox{in} \quad \mathbb{R}^3\setminus \overline{B},\\
 \nm\ds
\frac{1}{\mu_c}\bE^i(z)+\frac{\mu_m}{\mu_c^2-\mu_m\mu_c}\nabla\mathcal{S}_B\left(\lambda_\mu I - \mathcal{K}_B^*\right)^{-1}[{\nu}\cdot \bE^i(z)] & \mbox{in} \quad B,
\end{array}
\right.
\end{equation}
and
\begin{equation}\label{eq:lepsi}
\ds \nabla\times\mathcal{A}_B[\widetilde{{\psi}}_{0,0}]=\left\{
\begin{array}{ll}
 \frac{i}{\omega(\varepsilon_c-\varepsilon_m)}\nabla\mathcal{S}_B\left(\lambda_\varepsilon I - \mathcal{K}_B^*\right)^{-1}[{\nu}\cdot \bH^i(z)] & \mbox{in} \quad \mathbb{R}^3\setminus \overline{B} ,\\
 \nm \ds
\frac{i}{\omega\varepsilon_c}\bH^i(z)+\frac{i\varepsilon_m}{\omega(\varepsilon_c^2-\varepsilon_m\varepsilon_c)}\nabla\mathcal{S}_B\left(\lambda_\varepsilon I - \mathcal{K}_B^*\right)^{-1}[{\nu}\cdot \bH^i(z)] & \mbox{in} \quad B .
\end{array}
\right.
\end{equation}
\end{lemm}

\begin{proof}
 We only prove (\ref{eq:lephi}). We shall consider the solution to the following system
\begin{equation}\label{eq:letmp}
\left\{
\begin{array}{ll}
\Delta u=0 & \mbox{in}\quad \mathbb{R}^3, \\
(\nubf\cdot \nabla u )^- = (\nubf\cdot \nabla u)^+ & \mbox{on}\quad \dr B, \\
\mu_c (\nu\times \nabla u )^- -\mu_m (\nu\times\nabla u)^+=\nubf\times \bE^i(z) & \mbox{on}\quad \dr B, \\
u=O(|x|^{-1}) & |x|\rightarrow \infty.
\end{array}
\right.
\end{equation}

We can see that both the left-hand side and the right-hand side of (\ref{eq:lephi}) are  divergence free. We want to prove that they are both equal to the field $\nabla u$ in $\mathbb{R}^3$.
First we check that they satisfy the jump relations.
We already have the continuity of the normal part of the curl of a vectorial single layer potential \cite{colton2013integral}.
Recall that
\begin{equation*}
\widetilde{\phibf}_{0,0}=(\mu_c-\mu_m)^{-1}(\lambda_{\mu}I - \mathcal{M}_B)^{-1}[\nubf(\widetilde{y})\times \bE^i(z)].
\end{equation*}
Then,
\begin{equation*}
 \left( \nubf\times \g \times \mathcal{A}_D[\widetilde{\phibf}_{0,0}] 
 \right)^\pm=\frac{1}{\mu_c-\mu_m} \left(\mp\frac{I}{2} + \mathcal{M}_B\right) \left( \lambda I - \mathcal{M}_B\right)^{-1} [ \nubf(\widetilde{y})\times \bE^i(z)],
\end{equation*} so we have
\begin{equation*}
\mu_c  \big( \nubf\times \g \times \mathcal{A}_D[\widetilde{\phibf}_{0,0}] \big)^- - \mu_m  \bigr(\nubf\times \g \times \mathcal{A}_D[\widetilde{\phibf}_{0,0}] \big)^+ = \nubf(\widetilde{y})\times \bE^i(z).
\end{equation*}
The continuity of the tangential derivative of a scalar single layer potential gives 
\begin{multline*}
\mu_c \big( \nubf \times \left( \frac{1}{\mu_c-\mu_m}\nabla\mathcal{S}_B\left(\lambda_\mu I - \mathcal{K}_B^*\right)^{-1}[{\nu}\cdot \bE^i(z)]\big)^-\right) = \mu_m \big( \nubf \times \bigg( \frac{1}{\mu_c}\bE^i(z)\\ +\frac{\mu_m}{\mu_c^2-\mu_m\mu_c}\nabla\mathcal{S}_B\left(\lambda_\mu I -  \mathcal{K}_B^*\right)^{-1}[{\nu}\cdot \bE^i(z)]\big)^+ \bigg),
\end{multline*}
and the jump of the normal derivative of a scalar single layer potential can be written as follows
\begin{equation*}
\left( \nubf\cdot\g \mathcal{S}_B\left(\lambda_\mu I - \mathcal{K}_B^*\right)^{-1}[{\nu}\cdot \bE^i(z)]\right)^\pm =\left( \mp \frac{I}{2} + \mathcal{K}_B^*\right) \left(\lambda_\mu I - \mathcal{K}_B^*\right)^{-1}[{\nu}\cdot \bE^i(z)],
\end{equation*} which gives the correct jump relation for the normal derivative.

 The only problem left is to prove the uniqueness of the system. Now let $\widetilde{u}$ be the solution to (\ref{eq:letmp}) with the term $\nubf\times \bE^i(z)$ replaced by vector $0$ on $\dr B$. Note that
$\mu_c (\nu\times \nabla \widetilde{u})^- =\mu_m (\nu\times\nabla \widetilde{u})^+$ is equivalent to
\begin{equation*}
\mu_c \big(\frac{\dr \widetilde{u}}{\dr T} \big)^- =\mu_m \big(\frac{\dr \widetilde{u}}{\dr T}\big)^+,
\end{equation*}
where $T$ is any tangential direction on $\dr B$. Then by choosing any test function in $H^1(\dr B)$ and integrating by parts we can get $\mu_c (\widetilde{u})^-=\mu_m (\widetilde{u})^+$ on $\dr B$. Thus,
\begin{equation*}
 0\leq \int_{\mathbb{R}^3} \mu|\nabla \widetilde{u}|^2 dx = -\int_{\dr B} \mu_m \big(\frac{\dr \widetilde{u}}{\dr \nubf}\big)^+ (\widetilde{u})^+
+\int_{\dr B} \mu_c \big(\frac{\dr \widetilde{u}}{\dr \nubf}\-big)^- (\widetilde{u})^-=0,
\end{equation*}
which proves $\widetilde{u}=0$ and completes the proof.\end{proof}

It is worth mentioning that  it was proved in \cite{griesmaier2008asymptotic} that
$$
\nabla\times\mathcal{A}_B(\frac{1}{2}I+\mathcal{M}_B)^{-1}[\nu\times \bE^i(z)]=\bE^i(z) \quad \mbox{in} \quad B,
$$
which, by taking $\mu_m=0$ (or let $\mu_c=\infty$), can be seen as the extreme case in (\ref{eq:lephi}).

Now that we have a better understanding of $\nubf\times \g \times \mathcal{A}_D[\widetilde{\phibf}_{0,0}]$, by Lemma \ref{le:phipsi}, we can introduce the unique solutions $u^e,u^h\in H^1(B)$ up to constants such that $\nabla u^e=\nabla\times\mathcal{A}_B[\widetilde{{\phi}}_{0,0}]$,
$\nabla u^h=\nabla\times\mathcal{A}_B[\widetilde{{\psi}}_{0,0}]$ with $u^e$, $u^h$ satisfying
\begin{equation}\label{eq:fst}
\left\{
\begin{array}{ll}
 \Delta u^e=0 & \mbox{in} \quad B,\\
 ({ \nu}\cdot \nabla u^e)^-= {\nu}\cdot (\nabla\times\mathcal{A}_B[\widetilde{{\phi}}_{0,0}]) & \mbox{on} \quad \dr B,
\end{array}
\right.
\end{equation}
and
\begin{equation}\label{eq:scd}
\left\{
\begin{array}{ll}
 \Delta u^h=0 & \mbox{in} \quad B,\\
 ({\nu}\cdot \nabla u^h)^-= {\nu}\cdot (\nabla\times\mathcal{A}_B[\widetilde{{\psi}}_{0,0}]) & \mbox{on} \quad \dr B.
\end{array}
\right.
\end{equation}
The expressions of $\nabla u^e$ and $\nabla u^h$ are given by Lemma \ref{le:phipsi}.
Now, by using equation (\ref{eq:divphi1psi1}), we can compute the surface divergence of $\widetilde{{\phi}}_{0,1}$ and $\widetilde{{\psi}}_{0,1}$:
\begin{align*}
&\nabla_{\dr B}\cdot\widetilde{{\phi}}_{0,1}=\frac{k_c^2-k_m^2}{\mu_c-\mu_m}(\lambda_\mu I-\mathcal{K}_B^*)^{-1}\Big[ \big(\frac{\dr u^h}{\dr {\nu}}\big)^-\Big],\\
&\nabla_{\dr B}\cdot\widetilde{{\psi}}_{0,1}=\frac{k_c^2-k_m^2}{\omega^2(\varepsilon_c-\varepsilon_m)}(\lambda_\varepsilon I-\mathcal{K}_B^*)^{-1}\Big[ \big(\frac{\dr u^e}{\dr {\nu}}\big)^-\Big].
\end{align*}
Then we have the following lemma.
\begin{lemm}\label{le:tns00}
Let $v^e$ be the solution to
\begin{equation}\label{eq:eh}
\left\{
\begin{array}{ll}
 \Delta v^e=0 & x\in \mathbb{R}^3\setminus \partial B,\\
 (v^e)^+- (v^e)^-=0 & x\in \dr B,\\
 \varepsilon_m \big(\frac{\dr v^e}{\dr {\nu}}\big)^+-\varepsilon_c \big(\frac{\dr v^e}{\dr {\nu}}\big)^-=(\varepsilon_c-\varepsilon_m) \big(\nabla u^e \cdot {\nu}\big)^- & x\in \dr B ,\\
 v^e\rightarrow 0 &  |x| \rightarrow \infty,
\end{array}
\right.
\end{equation}
and let $v^h$ be the solution to 
\begin{equation}\label{eq:vh}
\left\{
\begin{array}{ll}
 \Delta v^h=0 & x\in \mathbb{R}^3\setminus\dr B,\\
 (v^h)^+- (v^h)^-=0 & x\in \dr B,\\
 \mu_m \big(\frac{\dr v^h}{\dr {\nu}}\big)^+-\mu_c \big(\frac{\dr v^h}{\dr {\nu}}\big)^-=(\mu_c-\mu_m) \big(\nabla u^h \cdot {\nu}\big)^- & x\in \dr B ,\\
 v^h\rightarrow 0 &  |x| \rightarrow \infty.
\end{array}
\right.
\end{equation}
Then the following asymptotic expansions hold:
\begin{align*}
&\mathrm{M}^e_{0,0}= \delta \frac{k_m^2-k_c^2}{\omega^2\varepsilon_m}\int_B\nabla (u^e+ v^e) +O(\delta^2),\\
&\mathrm{M}^h_{0,0}=\delta \frac{k_m^2-k_c^2}{\mu_m}\int_B\nabla (u^h+ v^h) +O(\delta^2).
\end{align*}
\end{lemm}
\begin{proof} By Proposition \ref{estimatephi_beta_n}, we have
\begin{align*}
\mathrm{M}^{h}_{0,0}& =\int_{\dr B} \widetilde{{\phi}}_{0} d\sigma(\widetilde{y})= \delta \int_{\dr B} \widetilde{{\phi}}_{0,1} d\sigma(\widetilde{y}) +O(\delta^2)\\
&=-\delta \int_{\dr B} \widetilde{y}\nabla_{\dr B}\cdot\widetilde{{\phi}}_{0,1} d\sigma(\widetilde{y}) +O(\delta^2)\\
& =\delta \frac{k_m^2-k_c^2}{\mu_c-\mu_m}\int_{\dr B} \widetilde{y}(\lambda_\mu I- \mathcal{K}_B^*)^{-1}\Big[ \big(\frac{\dr u^h}{\dr {\nu}}\big)^-\Big]d\sigma(\widetilde{y}) +O(\delta^2).
\end{align*}
Using the fact that
\begin{equation*}
\lambda_\mu=\frac{1}{2}+\frac{\mu_m}{\mu_c-\mu_m},
\end{equation*}
we get that for $f\in L^2(\dr B)$, $$\ds f=\frac{\mu_c-\mu_m}{\mu_m}\left[ \left(\lambda_\mu I - \mathcal{K}_B^*\right)[f] + \left( -\frac{I}{2} + \mathcal{K}_D^*\right) [f]\right].$$ Then,
\begin{equation*}
\mathrm{M}^{h}_{0,0} =\delta \frac{k_m^2-k_c^2}{\mu_m}\Big(\int_{\dr B} \widetilde{y} \big(\frac{\dr u^h}{\dr {\nu}}\big)^-d\sigma(\widetilde{y}) +\int_{\dr B}\widetilde{y}(-\frac{I}{2}+\mathcal{K}_B^*)(\lambda_\mu I- \mathcal{K}_B^*)^{-1}\Big[ \big(\frac{\dr u^h}{\dr {\nu}}\big)^-\Big]d\sigma(\widetilde{y})\Big)+O(\delta^2).
\end{equation*}
An integration by parts gives $$\ds \int_{\dr B} \widetilde{y} \big(\frac{\dr u^h}{\dr {\nu}}\big)^- d\sigma(\widetilde{y}) = \int_B\nabla u^hdx.$$
We now take a look at the transmission problem (\ref{eq:vh}) solved by $v^h$. Using the jump relation of the normal derivative of the scalar single layer potential we find that, writing $\ds v^h=\mathcal{S}_B[f] $ with $f$ being such that $ \ds \left(\lambda_\mu I - \mathcal{K}^*\right) [f] =  \frac{\dr u^h}{\dr {\nu}}$ gives $$\ds \left(-\frac{I}{2} +\mathcal{K}_B^*\right) [f]= \big( \frac{\dr v^h}{\dr \nubf}\big)^-,$$ and hence,
\begin{equation*}
(-\frac{I}{2}+\mathcal{K}_B^*)(\lambda_\mu I- \mathcal{K}_B^*)^{-1}\Big[ \big(\frac{\dr u^h}{\dr {\nu}}\big)^-\Big] = \big(\frac{\dr v^h}{\dr \nubf}\big)^-. 
\end{equation*}
Integrating by parts we get
\begin{equation*}
\mathrm{M}^{h}_{0,0}=\delta \frac{k_m^2-k_c^2}{\mu_m}\Big(\int_B\nabla u^hdx
+\int_B\nabla v^h dx\Big)+O(\delta^2).
\end{equation*}
The evaluation for $\mathrm{M}^{e}_{0,0}$ can be done  in exactly the same way.
\end{proof}

\subsubsection{Derivation of the leading-order tensors}

\begin{lemm}\label{le:epan1st}
We have
\begin{align}
&\mathrm{M}^e_{\alpha,\beta}=\frac{i}{\omega\varepsilon_m}\Big(\int_B\nabla(x^{\alpha}x^{\beta})\times \dr^{\beta}\bH^i(z)
+i\omega(\varepsilon_c-\varepsilon_m)\int_B\nabla\times (x^{\alpha}\nabla \times \mathcal{A}_B[\widetilde{{\psi}}_{\beta,0}])\Big)+O(\delta),\label{eq:epan11}\\
&\mathrm{M}^h_{\alpha,\beta}=\frac{1}{\mu_m}\Big(\int_B\nabla(x^{\alpha}x^{\beta})\times \dr^{\beta}\bE^i(z)
-(\mu_c-\mu_m)\int_B\nabla\times (x^{\alpha}\nabla \times \mathcal{A}_B[\widetilde{{\phi}}_{\beta,0}])\Big) +O(\delta).\label{eq:epan12}
\end{align}
In particular, we have
\begin{align}
&\mathrm{M}^e_{j,0}=\frac{i}{\omega\varepsilon_m}|B|{e}_j\times \bH^i(z)-\frac{\varepsilon_c-\varepsilon_m}{\varepsilon_m} {e}_j\times\int_B  \nabla u^h + O(\delta) ,\\
&\mathrm{M}^h_{j,0}=\frac{1}{\mu_m}|B| {e}_j\times \bE^i(z)-\frac{\mu_c-\mu_m}{\mu_m}{e}_j\times \int_B \nabla u^e + O(\delta) ,\\
 \label{eq:Me0j} &\mathrm{M}^e_{0,j}=\frac{i}{\omega\varepsilon_m}|B|{e}_j\times \dr_j\bH^i(z) -\frac{\varepsilon_c-\varepsilon_m}{\varepsilon_m}\int_B \nabla\mathcal{S}_B[\nabla_{\dr B}\cdot \widetilde{{\psi}}_{j,0}] +O(\delta),\\
&\mathrm{M}^h_{0,j}=\frac{1}{\mu_m}|B|{e}_j\times \dr_j\bE^i(z)-\frac{\mu_c-\mu_m}{\mu_m}\int_B \nabla\mathcal{S}_B[\nabla_{\dr B}\cdot \widetilde{{\phi}}_{j,0}] +O(\delta), \label{eq:Mh0j}
\end{align}
where ${(e}_1,e_2,e_3)$ is an orthonormal basis of $\mathbb{R}^3$.
\end{lemm}
\begin{proof}  We shall only consider $\mathrm{M}^h_{\alpha,\beta}$.   
$\mathrm{M}^e_{\alpha,\beta}$ can be calculated in exactly the same way. We have
$$
\mathrm{M}^h_{\alpha,\beta}= (\mathrm{M}^{h}_{\alpha,\beta})^{(0)}+O(\delta),
$$
where  $(\mathrm{M}^{h}_{\alpha,\beta})^{(0)}$ is given by
$$
(\mathrm{M}^{h}_{\alpha,\beta})^{(0)}=\int_{\dr B} \widetilde{y}^{\alpha} \widetilde{{\phi}}_{\beta,0} d\sigma(\widetilde{y}).
$$
Since $\ds \lambda_\mu=\frac{1}{2}+\frac{\mu_m}{\mu_c-\mu_m}$ we have that for any ${f}\in L^2_T(\dr B)$,
\begin{equation*}
\left(\lambda_\mu I - \mathcal{M}_B\right) [{f}] -\left(\frac{I}{2}  + \mathcal{M}_B\right)[{f}] = \frac{\mu_m}{\mu_c-\mu_m} {f}.
\end{equation*}
By applying Proposition \ref{estimatephi_beta_n}, it follows that
\begin{multline*}
(\mathrm{M}^{h}_{\alpha,\beta})^{(0)} =\frac{1}{\mu_m}\int_{\dr B}  \widetilde{y}^{\alpha}{\nu}(\widetilde{y}) \times (\widetilde{y}^{\beta}\dr^{\beta} \bE^i(z))d\sigma(\widetilde{y})
\\-\frac{1}{\mu_m}\int_{\dr B}\widetilde{y}^{\alpha} (\frac{I}{2}+\mathcal{M}_B)(\lambda_{\mu}I - \mathcal{M}_B)^{-1}[{\nu}(\widetilde{y}) \times\widetilde{y}^{\beta}\dr^{\beta} \bE^i(z)]d\sigma(\widetilde{y}).
\end{multline*}
Using the jump relations on $\mathcal{M}_B$ and the fact that $$\ds \widetilde{\phibf}_{\beta,0} = \frac{1}{\mu_c-\mu_m}\left( \lambda_\mu I - \mathcal{M}_B\right)^{-1}[{ \nu}(\widetilde{y}) \times\widetilde{y}^{\beta}\dr^{\beta}],$$ we can write
\begin{multline*}
(\mathrm{M}^{h}_{\alpha,\beta})^{(0)} =\frac{1}{\mu_m}\int_{\dr B}  \widetilde{y}^{\alpha}{ \nu}(\widetilde{y}) \times (\widetilde{y}^{\beta}\dr^{\beta} \bE^i(z))d\sigma(\widetilde{y})\\-\frac{\mu_c-\mu_m}{\mu_m}\int_{\dr B}\widetilde{y}^{\alpha}{ \nu}(\widetilde{y})\times \nabla \times \big(\mathcal{S}_B[\widetilde{{\phi}}_{\beta,0}]\big)^- d\sigma(\widetilde{y}).
\end{multline*}
The curl theorem yields
\begin{equation*}
(\mathrm{M}^{h}_{\alpha,\beta})^{(0)}=\frac{1}{\mu_m}\int_B\nabla(x^{\alpha}x^{\beta})\times \dr^{\beta}\bE^i(z)dx
-\frac{\mu_c-\mu_m}{\mu_m}\int_B\nabla\times (x^{\alpha}\nabla \times \mathcal{S}_B[\widetilde{{\phi}}_{\beta,0}])dx,
\end{equation*}
and thus (\ref{eq:epan12})  holds. By using the definition of $u^e$ and $u^h$ we get the  case where $|\alpha|=1$, $|\beta|=0$.\end{proof}

\subsubsection{Derivation of the polarization tensor}

Denote by ${G}(x,z)$ the matrix valued function (Dyadic Green function)
\begin{equation*}
{G}(x,z)=\varepsilon_m(\Gamma^{k_m}(x-z) {I} +  \frac{1}{k_m^2}D_{x}^2\Gamma^{k_m}(x-z)).
\end{equation*}
It can be seen that ${G}(x,z)$ satisfies
\begin{equation*}
\nabla_{x}\times\frac{1}{\varepsilon_m}\nabla_{x}\times {G}(x,z) -\omega^2\mu_m {G}(x,z)= -\delta_{z} {I}.
\end{equation*}
We can also easily check that
\begin{equation*}
\nabla\times {G}(x,z)=\varepsilon_m\nabla\times (\Gamma^{k_m}(x-z) {I})= \varepsilon_m\nabla \Gamma^{k_m}(x-z)\times {I}.
\end{equation*}
\begin{theo} \label{theo:expansion}
Define the polarization tensors
\begin{equation} \label{defmemh}
{M}^e:=\int_{\dr B} \widetilde{y} (\lambda_{\varepsilon}I-\mathcal{K}_B^*)^{-1}[\nu] d\sigma(\widetilde{y}) \quad \mbox{and}\quad
{M}^h:=\int_{\dr B} \widetilde{y} (\lambda_{\mu}I-\mathcal{K}_B^*)^{-1}[\nu] d\sigma(\widetilde{y}).
\end{equation}
Then the following far-field expansion holds:

\begin{equation}
\bE(x)-\bE^i(x) =-\delta^3\omega^2\mu_m {G}(x,z) {M}^e\bE^i(z)-\delta^3\frac{i\omega\mu_m}{\varepsilon_m}\nabla\times {G}(x,z) {M}^h\bH^i(z) + O(\delta^4).
\end{equation}
\end{theo}

Before we proceed, we stress that the polarization tensors ${M}^e$, ${M}^h$ defined above are matrix with each entry ${m}_{ij}^e$ and 
${m}_{ij}^h$, $i,j=1,2,3$, defined by (\ref{polarizationtensionr}) with $\lambda= \lambda_\varepsilon$ and 
$\lambda= \lambda_\mu$, respectively. 

%$$
%{M}_{ij}^e:=\int_{\dr B} \widetilde{y}_i (\lambda_{\varepsilon}I-\mathcal{K}_B^*)^{-1}[\nu_j] d\sigma(\widetilde{y}) \quad \mbox{and}\quad
%{M}_{ij}^h:=\int_{\dr B} \widetilde{y}_i (\lambda_{\mu}I-\mathcal{K}_B^*)^{-1}[\nu_j] d\sigma(\widetilde{y}).
%$$
They are different from the vector valued tensors we defined in equation (\ref{eq:tns}).

\begin{proof}
We shall give the analysis term by term in (\ref{eq:expansion01}). It is easy to check that
$$
\sum_{j=1}^3 {e}_j\times \dr_j\bE^i(z)=i\omega\mu_m\bH^i(z)\quad \mbox{and}\quad \sum_{j=1}^3 {e}_j\times \dr_j\bH^i(z)=-i\omega\varepsilon_m\bE^i(z)
$$
and
$$
\sum_{j=1}^3\nabla\dr_j\Gamma^{k_m}(x-z)\times {e}_j\times \bE^i(z)=\omega^2\mu_m {G}(x,z)\bE^i(z).
$$
Then by Lemma \ref{le:epan1st} it follows that
\begin{multline*}
 \nabla\times\sum_{j=1}^{3}\nabla\dr_{j}\Gamma^{k_m}(x-z)\times \mathrm{M}^e_{j,0} =\\
  \omega^2\mu_m\nabla\times {G}(x,z)\Big(\frac{i}{\omega\varepsilon_m}|B|\bH^i(z)-\frac{\varepsilon_c-\varepsilon_m}{\varepsilon_m}\int_B \nabla u^h \Big)+O(\delta),
\end{multline*}
and
\begin{multline*}
\mu_m\sum_{j=1}^{3}\nabla\dr_{j}\Gamma^{k_m}(x-z)\times \mathrm{M}^h_{j,0}=
 \omega^2\mu_m {G}(x,z)\Big(|B|\bE^i(z)-(\mu_c-\mu_m)\int_B \nabla u^e  \Big)\\+O(\delta).
\end{multline*}
Furthermore, we obtain from Proposition \ref{estimatephi_beta_n} that
\begin{align*}
\sum_{j=1}^3\nabla_{\dr B}\cdot \widetilde{{\phi}}_{j,0}=& \frac{1}{\mu_c-\mu_m} \left(\lambda_\mu I -\mathcal{K}_B^*\right)^{-1}\left[\sum_{j=1}^3 \g_{\dr B} \cdot \left(\nubf(\widetilde{y}) \times (\widetilde{y}_j \dr_j\bE^i(z))\right)\right],\\
\sum_{j=1}^3\nabla_{\dr B}\cdot \widetilde{{\phi}}_{j,0}=& \frac{1}{\mu_c-\mu_m} \left(\lambda_\mu I -\mathcal{K}_B^*\right)^{-1}\left[\sum_{j=1}^3 \nubf(\widetilde{y}) \cdot \left( \g \times (\widetilde{y}_j \dr_j\bE^i(z))\right)\right],
\end{align*} which gives
\begin{equation*}
\sum_{j=1}^3\nabla_{\dr B}\cdot \widetilde{{\phi}}_{j,0}=-\frac{i\omega\mu_m}{\mu_c-\mu_m} \left(\lambda_\mu I -\mathcal{K}_B^*\right)^{-1} [\nu\cdot \bH^i(z)].
\end{equation*}
Similarly, we have
\begin{equation*}
\sum_{j=1}^3\nabla_{\dr B}\cdot \widetilde{{\psi}}_{j,0}= -\frac{\varepsilon_m}{\varepsilon_c-\varepsilon_m} \left(\lambda_\mu I -\mathcal{K}_B^*\right)^{-1} [\nu\cdot \bE^i(z)].
\end{equation*}
Recall from (\ref{eq:Me0j}) that 
\begin{equation*}
\mathrm{M}^e_{0,j}=\frac{i}{\omega\varepsilon_m}|B|{e}_j\times \dr_j\bH^i(z) -\frac{\varepsilon_c-\varepsilon_m}{\varepsilon_m}\int_B \nabla\mathcal{S}_B[\nabla_{\dr B}\cdot \widetilde{{\psi}}_{j,0}] +O(\delta).
\end{equation*}
Summing over $j$ gives
\begin{equation*}
\begin{aligned}
\nabla\times\nabla\Gamma^{k_m}(x-z)\times \sum_{j=1}^{3} \frac{i}{\omega\varepsilon_m}|B|{e}_j\times \dr_j\bH^i(z)=&\g\times \nabla\Gamma^{k_m}(x-z)\times\left( \frac{i}{\omega\varepsilon_m}|B| \g_z \times \bH^i(z)\right)\\
=& -\g\times \nabla\Gamma^{k_m}(x-z)\times \vert B\vert \bE^i(z)\\
=& -\g \times \g  \times {G}(x,z) \vert B\vert \bE^i(z)\\
=& \omega^2\mu_m {G}(x,z) \vert B\vert \bE^i(z).
\end{aligned}
\end{equation*}
Hence, we can deduce that
\begin{equation*}
\nabla\times\nabla\Gamma^{k_m}(x-z)\times \sum_{j=1}^{3} \mathrm{M}^e_{0,j} =\omega^2 \mu_m {G}(x,z) \left(\vert B\vert \bE^i(z) + \int_B \nabla\mathcal{S}_B[\nubf\cdot\bH^i(z)]\right) + O(\delta).
\end{equation*}
 A similar computation yields
 \begin{multline*}
\mu_m\nabla\Gamma^{k_m}(x-z)\times\sum_{j=1}^{3}\mathrm{M}^h_{0,j}=\\
i\omega\mu_m\nabla\Gamma^{k_m}(x-z)\times\Big(|B|\bH^i(z)+\int_B \nabla\mathcal{S}_B(\lambda_\mu I-\mathcal{K}_B^*)^{-1}[\nu\cdot \bH^i(z)]\Big) +O(\delta),
\end{multline*}
and therefore, 
\begin{multline*}
\mu_m\nabla\Gamma^{k_m}(x-z)\times\sum_{j=1}^{3}\mathrm{M}^h_{0,j}=\\
i\omega\frac{\mu_m}{\varepsilon_m}\nabla\times {G}(x-z)\Big(|B|\bH^i(z)+\int_B \nabla\mathcal{S}_B(\lambda_\mu I-\mathcal{K}_B^*)^{-1}[\nu\cdot \bH^i(z)]\Big) +O(\delta).
\end{multline*}
Moreover, Lemma \ref{le:tns00} gives
\begin{align*}
\nabla\times\nabla\Gamma^{k_m}(x-z)\times \mathrm{M}^e_{0,0}=&\delta \frac{\mu_m}{\varepsilon_m}(k_m^2-k_c^2){G}(x,z)\int_B\nabla (u^e+v^e)+O(\delta^2)\\
\mu_m\nabla\Gamma^{k_m}(x-z)\times\mathrm{M}^h_{0,0}=&\delta\frac{(k_m^2-k_c^2)}{\varepsilon_m}\nabla\times {G}(x,z)\int_B\nabla (u^h+v^h)+O(\delta^2).
\end{align*}
Combining the previous asymptotic expansions we arrive at
\begin{multline}
\bE(x)-\bE^i(x) =\delta^3\frac{1}{\varepsilon_m} {G}(x,z)\Big(\mu_m(k_m^2-k_c^2)\int_B\nabla (u^e+v^e)\\+ (\mu_c-\mu_m)k_m^2\int_B\nabla u^e
\quad +k_m^2\int_B \nabla\mathcal{S}_B(\lambda_\varepsilon I-\mathcal{K}_B^*)^{-1}[\nu\cdot \bE^i(z)]\Big) \\
\quad + \delta^3\frac{1}{\varepsilon_m}\nabla\times {G}(x,z)\Big((k_m^2-k_c^2)\int_B\nabla (u^h+v^h)+ \omega^2\mu_m(\varepsilon_c-\varepsilon_m)\int_B\nabla u^h\\
\quad + i\omega\mu_m \int_B \nabla\mathcal{S}_B(\lambda_\mu I-\mathcal{K}_B^*)^{-1}[\nu\cdot \bH^i(z)]\Big)+O(\delta^4).\label{eq:epanmid}
\end{multline}
The proof is then complete. \end{proof}

We shall analyze further (\ref{eq:epanmid}). Recall that, from the proof of Lemma \ref{le:tns00},  we have
\begin{equation*}
\int_B \nabla (u^e+v^e)dx=\frac{\varepsilon_m}{\varepsilon_c-\varepsilon_m}\int_{\dr B} \widetilde{y} (\lambda_{\varepsilon}I-\mathcal{K}_B^*)^{-1}\Big[ \big(\frac{\dr u^e}{\dr \nu}\big)^-\Big]
 d\sigma(x)
\end{equation*}
and 
\begin{equation*}
\int_B \nabla (u^h+v^h)dx=\frac{\mu_m}{\mu_c-\mu_m}\int_{\dr B} \widetilde{y} (\lambda_{\mu}I-\mathcal{K}_B^*)^{-1}\Big[
\big(\frac{\dr u^h}{\dr \nu}\big)^-\Big]d\sigma(x).
\end{equation*}
Noticing that $$\mu_m(k_m^2-k_c^2)=(\mu_c-\mu_m)k_m^2\frac{\mu_m\varepsilon_m - \mu_c \varepsilon_c}{(\mu_c-\mu_m)(\varepsilon_c-\varepsilon_m)},$$ we get
\begin{multline*}
\mu_m(k_m^2-k_c^2)\int_B\nabla (u^e+v^e)+ (\mu_c-\mu_m)k_m^2\int_B\nabla u^e= \\
(\mu_c-\mu_m)k_m^2\left(\frac{\mu_m\varepsilon_m-\mu_c\varepsilon_c}{(\mu_c-\mu_m)(\varepsilon_c-\varepsilon_m)}\int_{\dr B} \widetilde{y}
(\lambda_{\varepsilon}I-\mathcal{K}_B^*)^{-1}\Big[\big(\frac{\dr u^e}{\dr \nu}\big)^-\Big]+\int_{\dr B} \widetilde{y} \big(\frac{\dr u^e}{\dr \nu}\big)^-\right).
\end{multline*}
Moreover, for any $f$, we have
\begin{multline*}
\frac{\mu_m\varepsilon_m-\mu_c\varepsilon_c}{(\mu_c-\mu_m)(\varepsilon_c-\varepsilon_m)} \left( \lambda_\varepsilon I -\mathcal{K}_B^*\right)^{-1} [f ] + f =\\  \frac{\mu_m\varepsilon_m-\mu_c\varepsilon_c}{(\mu_c-\mu_m)(\varepsilon_c-\varepsilon_m)} \left( \lambda_\varepsilon I -\mathcal{K}_B^*\right)^{-1} [f ] + \left( \lambda_\varepsilon I -\mathcal{K}^*\right)\left( \lambda_\varepsilon I -\mathcal{K}_B^*\right) ^{-1} [f],
\end{multline*} 
so that
\begin{equation*}
\frac{\mu_m\varepsilon_m-\mu_c\varepsilon_c}{(\mu_c-\mu_m)(\varepsilon_c-\varepsilon_m)} \left( \lambda_\varepsilon I -\mathcal{K}_B^*\right)^{-1} [f ] + f = (\lambda_\mu I+\mathcal{K}_B^*)(\lambda_{\varepsilon}I-\mathcal{K}_B^*)^{-1}[f]
\end{equation*} 
We can then write
\begin{multline*}
\mu_m(k_m^2-k_c^2)\int_B\nabla (u^e+v^e)+ (\mu_c-\mu_m)k_m^2\int_B\nabla u^e= \\
-(\mu_c-\mu_m)k_m^2\int_{\dr B}\widetilde{y} (\lambda_\mu I+\mathcal{K}_B^*)(\lambda_{\varepsilon}I-\mathcal{K}_B^*)^{-1}\Big[ \big(\frac{\dr u^e}{\dr \nu}\big)^-\Big].
\end{multline*}
Recall that by definition,
 \begin{equation*}
 \big(\frac{\dr u^e}{\dr \nubf} \big)^- = \nubf \cdot \g \times \mathcal{A}_B [\widetilde{\phibf}_{0,0}].
\end{equation*}
Then, by using Lemma \ref{le:phipsi}, we obtain
\begin{equation*}
\nubf \cdot \g \times \mathcal{A}_B [\widetilde{\phibf}_{0,0}]= \frac{1}{\mu_c}\nubf \cdot \bE^i(z)+\frac{\mu_m}{\mu_c^2-\mu_m\mu_c}
\big( \nubf\cdot\nabla\mathcal{S}_B\left(\lambda_\mu I - \mathcal{K}_B^*\right)^{-1}[{\nu}\cdot \bE^i(z)] \big)^-,
\end{equation*}
which together with the jump relations for the normal derivative of the scalar layer potential yields
\begin{multline*}
\mu_m(k_m^2-k_c^2)\int_B\nabla (u^e+v^e)+ (\mu_c-\mu_m)k_m^2\int_B\nabla u^e= \\
-\frac{\mu_c-\mu_m}{\mu_c}k_m^2\int_{\partial B}\widetilde{y} (\lambda_\mu I+\mathcal{K}_B^*)(\lambda_{\varepsilon}I-\mathcal{K}_B^*)^{-1}[\nu\cdot\bE^i(z)]\\
-\frac{\mu_m}{\mu_c}k_m^2\int_{\dr B}\widetilde{y} (\lambda_\mu I+\mathcal{K}_B^*)(\lambda_{\varepsilon}I-\mathcal{K}_B^*)^{-1}(-\frac{1}{2}I+\mathcal{K}_B^*)(\lambda_\mu I -\mathcal{K}_B^*)^{-1}[\nu\cdot \bE^i(z)].
\end{multline*}
If we set $\ds \lambda_\varepsilon=-\frac{1}{2}+\frac{\varepsilon_c}{\varepsilon_c-\varepsilon_m}$, then we can write
\begin{equation*}
-\frac{\mu_m}{\mu_c}\left(\lambda_\varepsilon I - \mathcal{K}_B^*\right)^{-1} \left(-\frac{1}{2} I + \mathcal{K}_B^*\right)[\nubf\cdot \bE^i(z)] = -\frac{\mu_m}{\mu_c}\left(\lambda_\varepsilon I - \mathcal{K}_B^*\right)^{-1} \left(\lambda_\varepsilon -\frac{\varepsilon_c}{\varepsilon_c - \varepsilon_m} I + \mathcal{K}_B^*\right)[\nubf\cdot \bE^i(z)],
\end{equation*}
or equivalently, 
\begin{multline*}
-\frac{\mu_m}{\mu_c}\left(\lambda_\varepsilon I - \mathcal{K}_B^*\right)^{-1} \left(-\frac{1}{2} I + \mathcal{K}_B^*\right)[\nubf\cdot \bE^i(z)] =\\ -\frac{\mu_m}{\mu_c} \nubf\cdot \bE^i(z) + \frac{\varepsilon_c\mu_m}{\mu_c (\varepsilon_c- \varepsilon_m)} \left(\lambda_\varepsilon I - \mathcal{K}_B^*\right)^{-1} [\nubf\cdot \bE^i(z)].
\end{multline*}
Then, since
\begin{multline*}
-\frac{\mu_c-\mu_m}{\mu_c} \left(\lambda_\mu I + \mathcal{K}_B^*\right)\left(\lambda_\varepsilon I - \mathcal{K}_B^*\right)^{-1}[\nubf\cdot \bE^i(z)]=\\ \frac{\mu_c-\mu_m}{\mu_c} \nubf\cdot \bE^i(z)- \frac{\mu_c-\mu_m}{\mu_c} (\lambda_\mu - \lambda_\varepsilon) \left(\lambda_\varepsilon I - \mathcal{K}_B^*\right)^{-1}[\nubf\cdot \bE^i(z)],
\end{multline*}
we can write
\begin{multline*}
-\frac{\mu_c-\mu_m}{\mu_c} \left(\lambda_\mu I + \mathcal{K}_B^*\right)\left(\lambda_\varepsilon I - \mathcal{K}_B^*\right)^{-1}[\nubf\cdot \bE^i(z)] \\-\frac{\mu_m}{\mu_c}\left(\lambda_\varepsilon I - \mathcal{K}_B^*\right)^{-1} \left(-\frac{1}{2} I + \mathcal{K}_B^*\right)[\nubf\cdot \bE^i(z)]  \\ = \nubf\cdot \bE^i(z) +\left(\frac{\varepsilon_c\mu_m}{\mu_c (\varepsilon_c- \varepsilon_m)} - \frac{\mu_c-\mu_m}{\mu_c} (\lambda_\mu - \lambda_\varepsilon) \right)\left(\lambda_\varepsilon I - \mathcal{K}_B^*\right)^{-1}[\nubf\cdot \bE^i(z)].
\end{multline*}
A direct computation gives 
\begin{equation*}
\frac{\varepsilon_c\mu_m}{\mu_c (\varepsilon_c- \varepsilon_m)} - \frac{\mu_c-\mu_m}{\mu_c} (\lambda_\mu - \lambda_\varepsilon) =\frac{1}{2}+ \lambda_\varepsilon,
\end{equation*}
and therefore, 
\begin{multline*}
\mu_m(k_m^2-k_c^2)\int_B\nabla (u^e+v^e)+ (\mu_c-\mu_m)k_m^2\int_B\nabla u^e= \\
k_m^2\int_{\dr B} \widetilde{y}\nubf\cdot \bE^i(z)d\sigma(\widetilde{y})  - k_m^2\left(\frac{1}{2}+ \lambda_\varepsilon\right) \int_{\dr B} \widetilde{y} \left(\lambda_\varepsilon I - \mathcal{K}_B^*\right)^{-1}[\nubf\cdot \bE^i(z)] d\sigma(\widetilde{y}).
\end{multline*}
A similar computation yields
\begin{multline*}
(k_m^2-k_c^2)\int_B\nabla (u^h+v^h)+\omega^2\mu_m (\varepsilon_c - \varepsilon_m) \int_B\nabla u^h= \\
i\omega\mu_m\int_{\dr B} \widetilde{y}\nubf\cdot \bH^i(z)d\sigma(\widetilde{y})  - i\omega\mu_m\left(\frac{1}{2}+ \lambda_\mu\right) \int_{\dr B} \widetilde{y} \left(\lambda_\varepsilon I - \mathcal{K}_B^*\right)^{-1}[\nubf\cdot \bH^i(z)] d\sigma(\widetilde{y}).
\end{multline*}
Now it remains to compute the last term in  (\ref{eq:epanmid}) which is
\begin{equation*}
k_m^2\int_B \nabla\mathcal{S}_B(\lambda_\varepsilon I-\mathcal{K}_B^*)^{-1}[\nu\cdot \bE^i(z)]d\widetilde{y} = k_m^2\int_{\dr B} \widetilde{y} \big(\frac{\dr}{\dr \nubf} \mathcal{S}_B\big)^-\left(\lambda_\varepsilon I - \mathcal{K}_B^*\right)^{-1} [\nubf \cdot \bE^i(z)]d\sigma(\widetilde{y}).
\end{equation*}
Writing that $\ds \lambda_\varepsilon = \frac{1}{2} + \frac{\varepsilon_m}{\varepsilon_c+\varepsilon_m}$ together with the fact that $\ds 
\big( \frac{\dr}{\dr \nubf} \mathcal{S}_B\big)^- = \left(-\frac{1}{2} I + \mathcal{K}_B^*\right)$, we obtain
\begin{equation*}
\frac{\dr}{\dr \nubf} \mathcal{S}_B \left(\lambda_\varepsilon I - \mathcal{K}_B^*\right)^{-1} [\nubf \cdot \bE^i(z)]= -\nubf\cdot \bE^i(z)  + \frac{\varepsilon_m}{\varepsilon_c+\varepsilon_m} \left(\lambda_\varepsilon I - \mathcal{K}_B^*\right)^{-1} [\nubf \cdot \bE^i(z)].
\end{equation*}
Hence, 
\begin{multline*}
k_m^2\int_B \nabla\mathcal{S}_B(\lambda_\varepsilon I-\mathcal{K}_B^*)^{-1}[\nu\cdot \bE^i(z)]d\widetilde{y} = -k_m^2\int_{\dr B} \widetilde{y} \nubf \cdot \bE^i(z)]d\sigma(\widetilde{y})\\ + k_m^2 \left(\lambda_\varepsilon - \frac{1}{2}\right)\int_{\dr B} \widetilde{y} \left(\lambda_\varepsilon I - \mathcal{K}_B^*\right)^{-1} [\nubf \cdot \bE^i(z)]d\sigma(\widetilde{y}).
\end{multline*}
Similarly, we have
\begin{multline*}
 i\omega\mu_m \int_B \nabla\mathcal{S}_B(\lambda_\mu I-\mathcal{K}_B^*)^{-1}[\nu\cdot \bH^i(z)] =i\omega\mu_m\int_{\dr B} \widetilde{y}\nubf\cdot \bH^i(z)d\sigma(\widetilde{y})   \\+i\omega\mu_m\left( \lambda_\mu-\frac{1}{2}\right) \int_{\dr B} \widetilde{y} \left(\lambda_\varepsilon I - \mathcal{K}_B^*\right)^{-1}[\nubf\cdot \bH^i(z)] d\sigma(\widetilde{y}).
\end{multline*}
Finally, we arrive at
\begin{multline*}
\bE(x)-\bE^i(x) =-\delta^3\omega^2\mu_m {G}(x,z)\int_{\partial B} \widetilde{y}(\lambda_{\varepsilon}I-\mathcal{K}_B^*)^{-1}[\nu\cdot\bE^i(z)] \\
 - \delta^3\frac{i\omega\mu_m}{\varepsilon_m}\nabla\times {G}(x,z)
\int_{\partial B} \widetilde{y}(\lambda_{\mu}I-\mathcal{K}_B^*)^{-1}[ \nu\cdot\bH^i(z)] + O(\delta^4).
\end{multline*}

When a plasmonic resonance occurs, the term $\lambda_\varepsilon = \frac{\varepsilon_c + \varepsilon_m}{2(\varepsilon_c-\varepsilon_m)}$ can have a real part that is lower than $\frac{1}{2}$, and become close to an eigenvalue of the operator $\mathcal{K}_B^*$. 

Let $d_\sigma$ be the minimum of the distances of $\lambda_\varepsilon$ and $\lambda_\mu$  to the spectrum $\sigma(\mathcal{K}^*_B)$.  Using Proposition \ref{estimatephi_beta_n} and (\ref{important2}) we can easily see that, as $d_\sigma$ goes to zero, each of the potentials $\phibf_{\beta,n}$ and $\psibf_{\beta,n}$ are controlled in norm by powers of $\frac{1}{d_\sigma}$. So the asymptotic development given by Theorem \ref{theo:expansion} is valid when  ${\delta}/{d_\sigma} <<1$, which ensures that the reminder of the asymptotic expansion is still small compared to the first-order term.

The following results are our main results in this paper. 
\begin{theo} \label{theo2f}
Let $d_\sigma:= \min \{\mathrm{dist}(\lambda_\varepsilon, \sigma(\mathcal{K}^*_B)), \mathrm{dist}(\lambda_\mu, \sigma(\mathcal{K}^*_B))\}$.
As $d_\sigma \rightarrow 0$, the following uniform far-field expansion holds:
\begin{equation*}
\begin{array}{lll}
\bE(x)-\bE^i(x) &=& \ds -\delta^3\omega^2\mu_m {G}(x,z) {M}^e\bE^i(z)-\delta^3\frac{i\omega\mu_m}{\varepsilon_m}\nabla\times {G}(x,z) {M}^h\bH^i(z) \\ && \ds + O(\frac{\delta^4}{d_\sigma} ),
\end{array}
\end{equation*}
where $M^e$ and $M^h$ are defined by (\ref{defmemh}). 
\end{theo}
The above theorem can be generalized to the case of multiple particles. 
\begin{theo} \label{theo3f}
Let ${M}^e$ and $M^h$ be the polarization tensors associated with $D_1 \cup D_2$ and  $\lambda_\varepsilon$ and $\lambda_\mu$, respectively. Let $d_\sigma:= \min \{\mathrm{dist}(\lambda_\varepsilon, \sigma(\mathbb{K}^*_B)), \mathrm{dist}(\lambda_\mu, \sigma(\mathbb{K}^*_B)) \}$.
Then as $d_\sigma \rightarrow 0$, the following uniform far-field expansion holds:
\begin{equation*}
\begin{array}{lll}
\bE(x)-\bE^i(x) &=&\ds -\delta^3\omega^2\mu_m {G}(x,z) {M}^e\bE^i(z)-\delta^3\frac{i\omega\mu_m}{\varepsilon_m}\nabla\times {G}(x,z) {M}^h\bH^i(z) \\ \nm && \ds + O(\frac{\delta^4}{d_\sigma}),
\end{array}
\end{equation*}
where $M^e$ and $M^h$ are defined by (\ref{polarizationtensionrm}) with $\lambda= \lambda_\varepsilon$ and $\lambda= \lambda_\mu$, respectively. 
\end{theo}

Theorems \ref{theo2f} and \ref{theo3f} show the uniform validity with respect to the nanoparticle's bulk electron relaxation rate of the quasi-static approximation of the Maxwell's equations.

Finally, two more remarks are in order. First, in view of Theorems \ref{theo2f} and \ref{theo3f} and the blow up of the polarization tensors, it is clear that at plasmonic resonances the scattered electric field is enhanced. Secondly, from the representation formula (\ref{represent}) for the electric field in $D$ and the estimates of the densities, it can be seen that the electric field inside the particle is enhanced as well and therefore,
the absorbed energy, given by $\varepsilon^{\prime\prime} \int_D |E|^2(y)\, dy$, is enhanced at dielectric plasmonic resonances \cite{matias}.  Note that the scattering enhancement when the particles are illuminated at their plasmonic resonances can be used for nano-resolved imaging from the far-field data while the absorption enhancement for thermotherapy applications as well as for photoacoustic imaging to remotely measure and control the local temperature within a medium \cite{fort}.

\section{Numerical illustrations} \label{sec23}
\begin{figure}

\begin{center}

\input{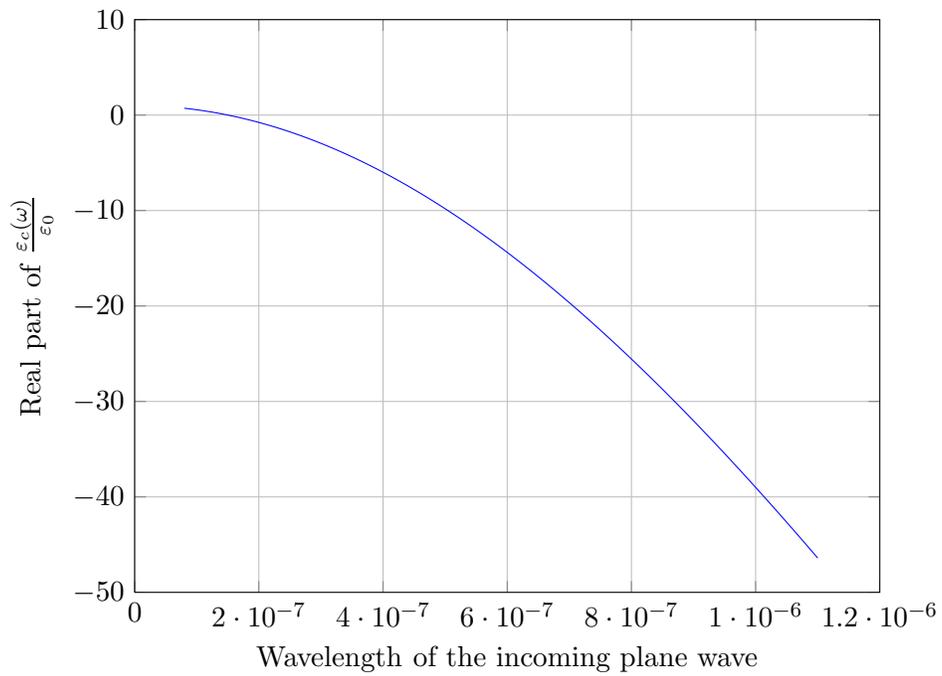}

\end{center}

\begin{center}

\input{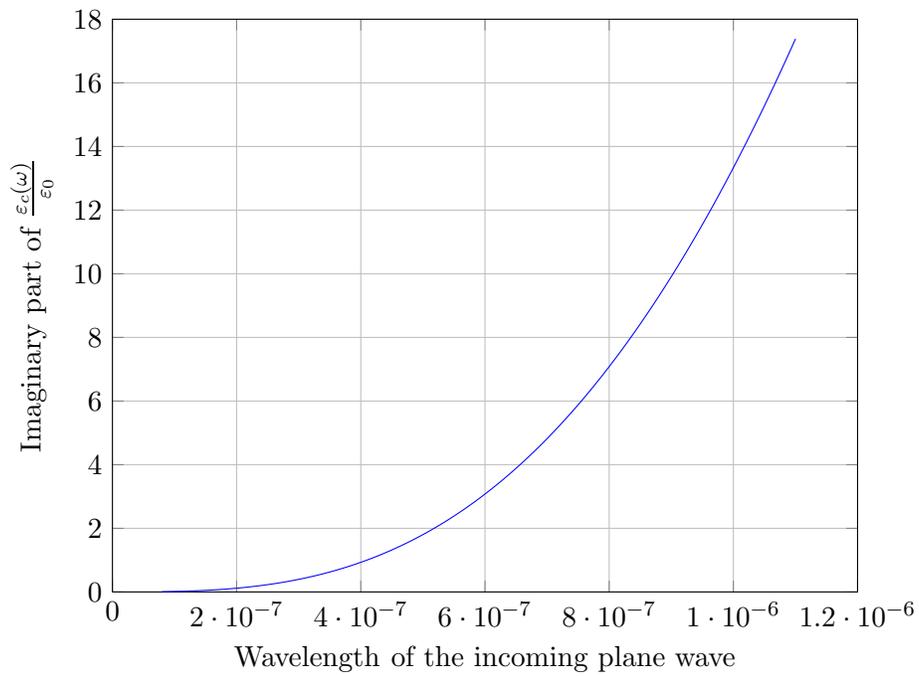}

\caption{\label{figajout1} Values of  the  parameter  $\varepsilon(\omega)$.}

\end{center}

\end{figure}

\begin{figure}

\begin{center}

\input{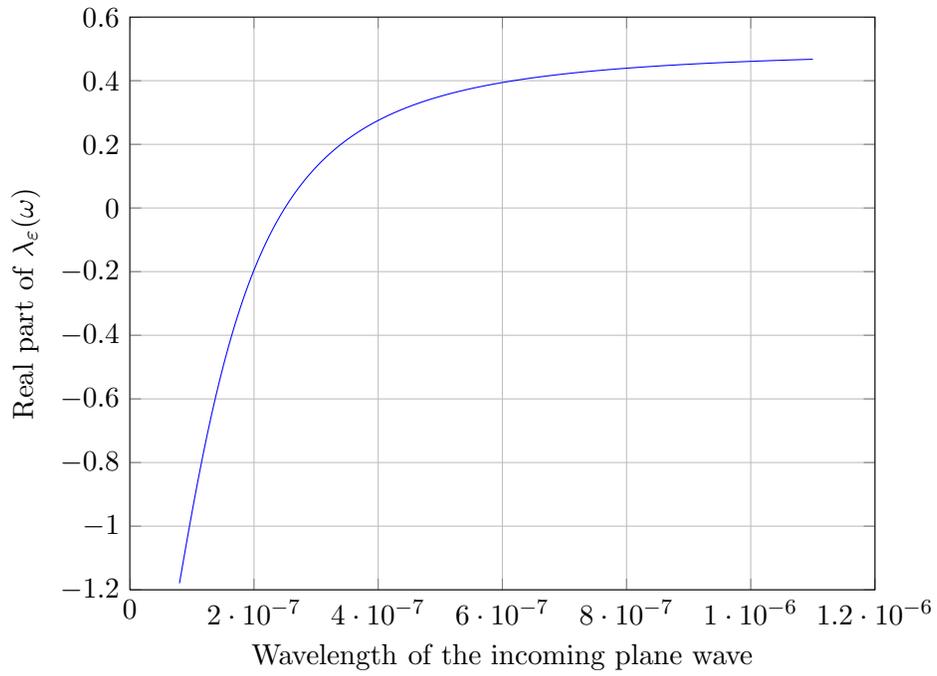}

\end{center}

\begin{center}

\input{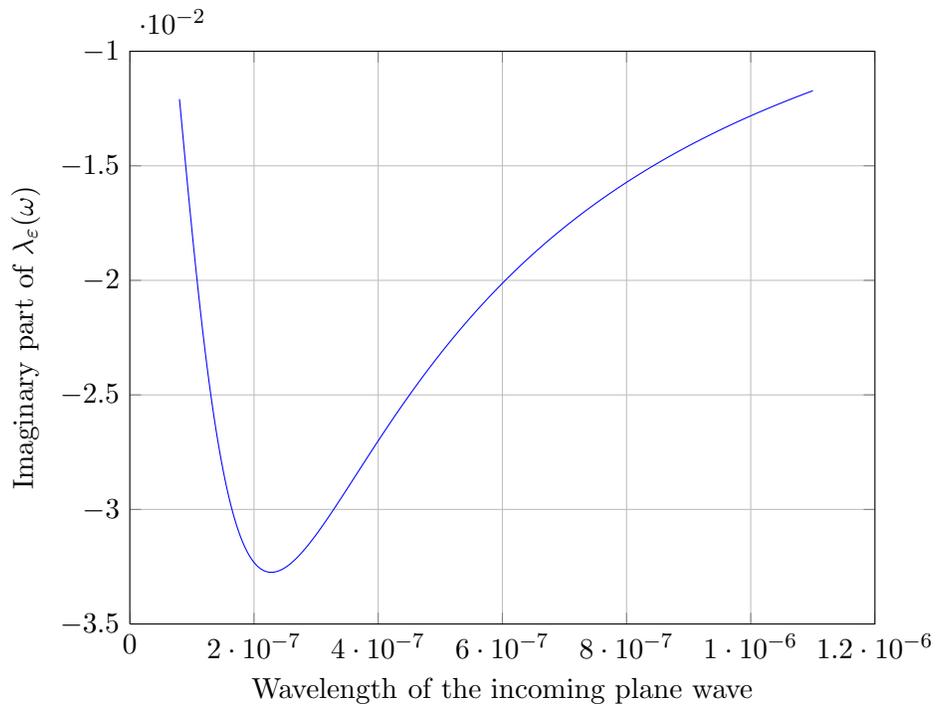}

\caption{\label{figajout2} Values of  the  parameter  $\lambda_\varepsilon(\omega)$.}

\end{center}

\end{figure}

We illustrate the plasmon phenomenon numerically by computing the polarization tensor 
${M}^e$ for some different two-dimensional shapes. We use the values for the parameters 
given in section \ref{sec23}. The wavelength of the incoming plane wave $c/\omega$, where $c= 3. 10^8$ is the speed of light, belongs to $[80, 1100] . 10^{-9} \, m.$  Figures \ref{figajout1} and \ref{figajout2} show respectively the values of real and imaginary parts of $\varepsilon_c$ and $\lambda_\varepsilon$ as a function of the wavelength.

%
%\begin{itemize}
%\item $\lambda \in [400 \cdot 10^{-9}; \ 4000 \cdot 10^{-9}]$ for the wavelength of the incoming plane wave,
%\item $c=3\cdot 10^8$  for the speed of light,
%\item $\varepsilon_0 = 9\cdot 10^{-12} F.m^{-1}$,
%\item $\omega_p = 2 \cdot 10^{15} s^{-1}$ ,
%\item $\tau = 10^{-14}s$,
%\item $\omega=\frac{2\pi c}{\lambda}$,
%\item $\varepsilon_m=2$,
%\item $\varepsilon_c(\omega)=\varepsilon_0(1-\frac{\omega_p^2}{\omega(\omega+\frac{i}{\tau})}) $.
% \end{itemize}

\begin{figure}
\begin{center}
\input{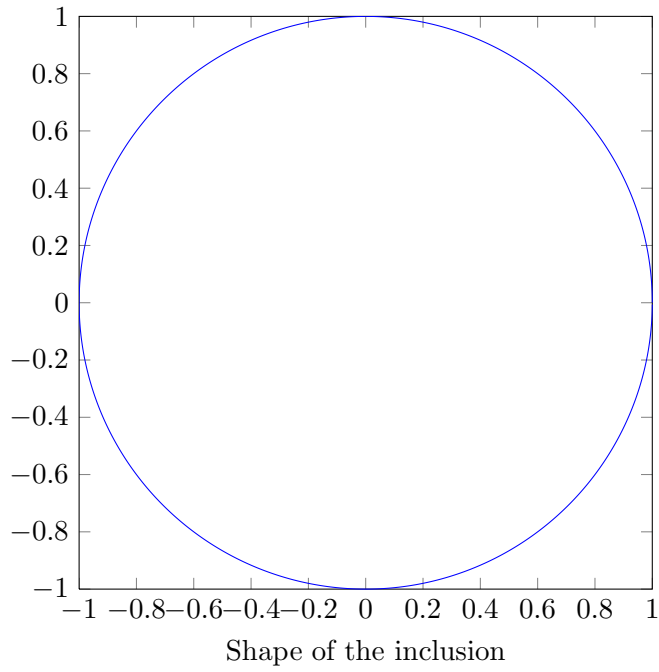}
\end{center}
\begin{center}
\input{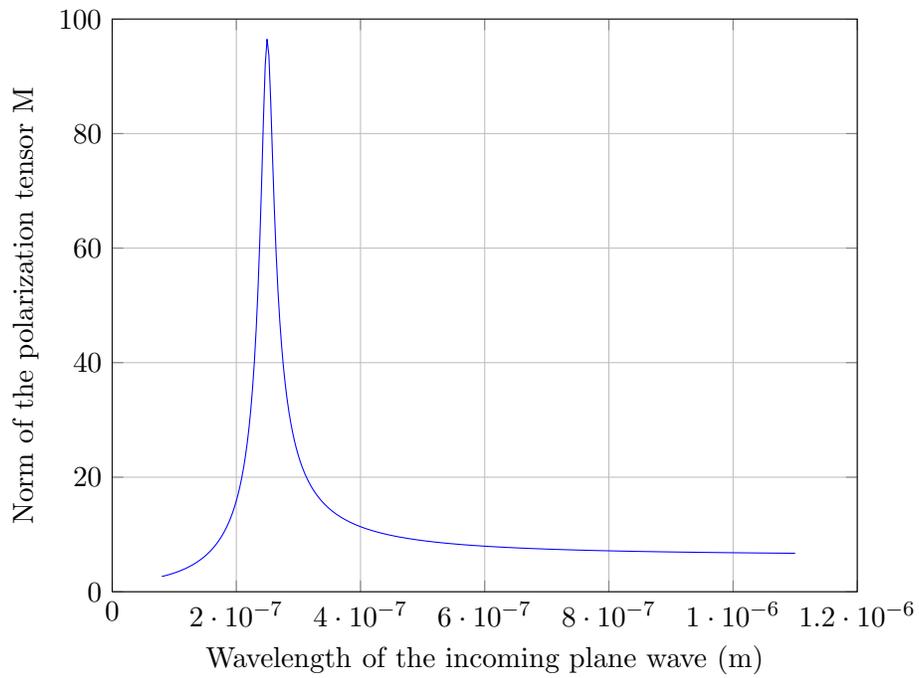}
\caption{\label{fig:circle} Norm of the polarization tensor for a circular inclusion.}
\end{center}
\end{figure}

\begin{figure}
\begin{center}
\input{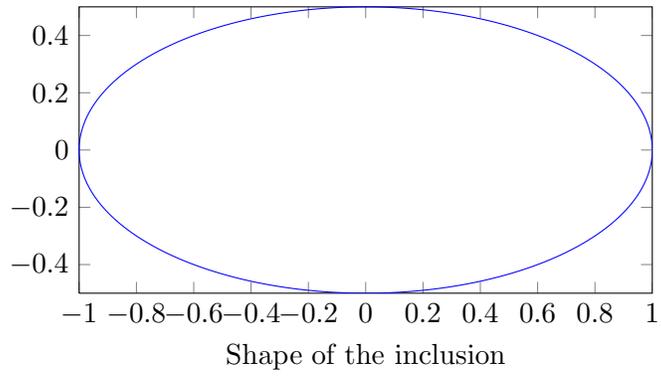}
\end{center}
\begin{center}
\input{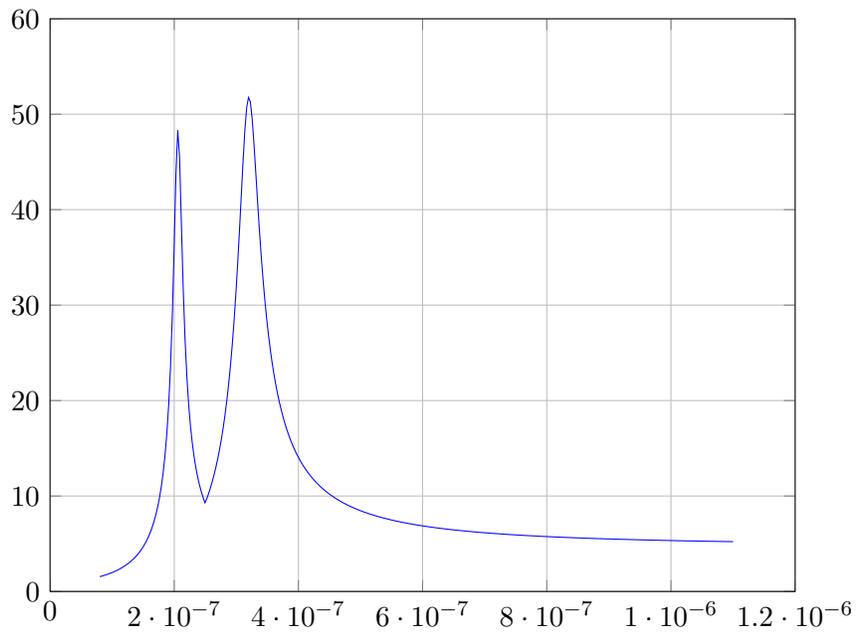}
\caption{\label{fig:ellipse} Norm of the polarization tensor for an elliptic inclusion.}
\end{center}
\end{figure}

\begin{figure}
\begin{center}
\input{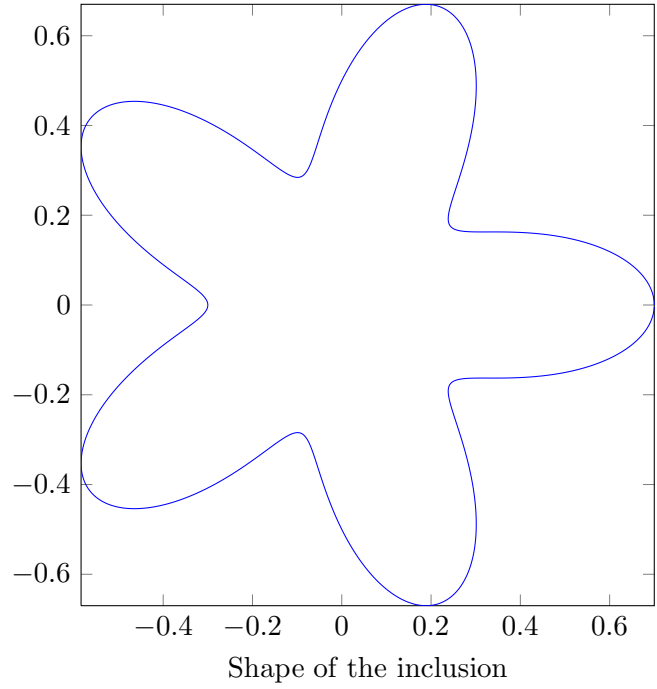}
\end{center}
\begin{center}
\input{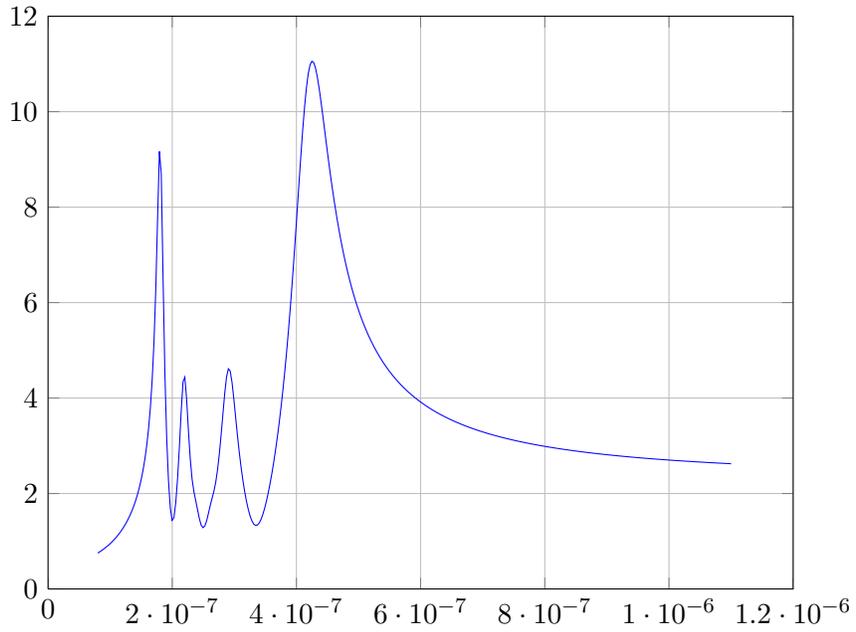}
\caption{\label{fig:flower} Norm of the polarization tensor for a flower-shaped inclusion.}
\end{center}
\end{figure}

Then we compute the matrix ${M}^e$ defined by (\ref{polarizationtensionr}) with $\lambda = \lambda_\varepsilon$. 
% 
% \begin{equation*}
% {M}^e_{i,j}=\int_{\dr B} \widetilde{y}_i (\lambda_{\varepsilon}I-\mathcal{K}_B^*)^{-1}[\nu_j] d\sigma(\widetilde{y}).
% \end{equation*}
 We plot the value of its norm with respect to the incoming wavelength. Figure \ref{fig:circle}
shows that if the shape $B$ is a disk, then one has a resonant peak. This peak corresponds to $\lambda_\varepsilon=0$. 
Figure \ref{fig:ellipse} shows that for an ellipse, one can observe two resonant frequencies, one corresponding to each axis. This was experimentally observed in \cite{el2008shape} for elongated particles. The two peaks correspond to $\lambda_\varepsilon =  (a-b)/(a+b)\approx 0.33$ and $\lambda_\varepsilon= ((a-b)/(a+b))^2\approx 0.11$, where $a=1,b=1/2$ are the semi-axis lengths of the ellipse. 
Figure \ref{fig:flower} gives the norm of the polarization tensor for a star-shaped particle. One can observe that there are many resonant frequencies. This observation is also in agreement with the experimental results published in \cite{hao2007plasmon}.

\begin{figure}
\begin{center}
\input{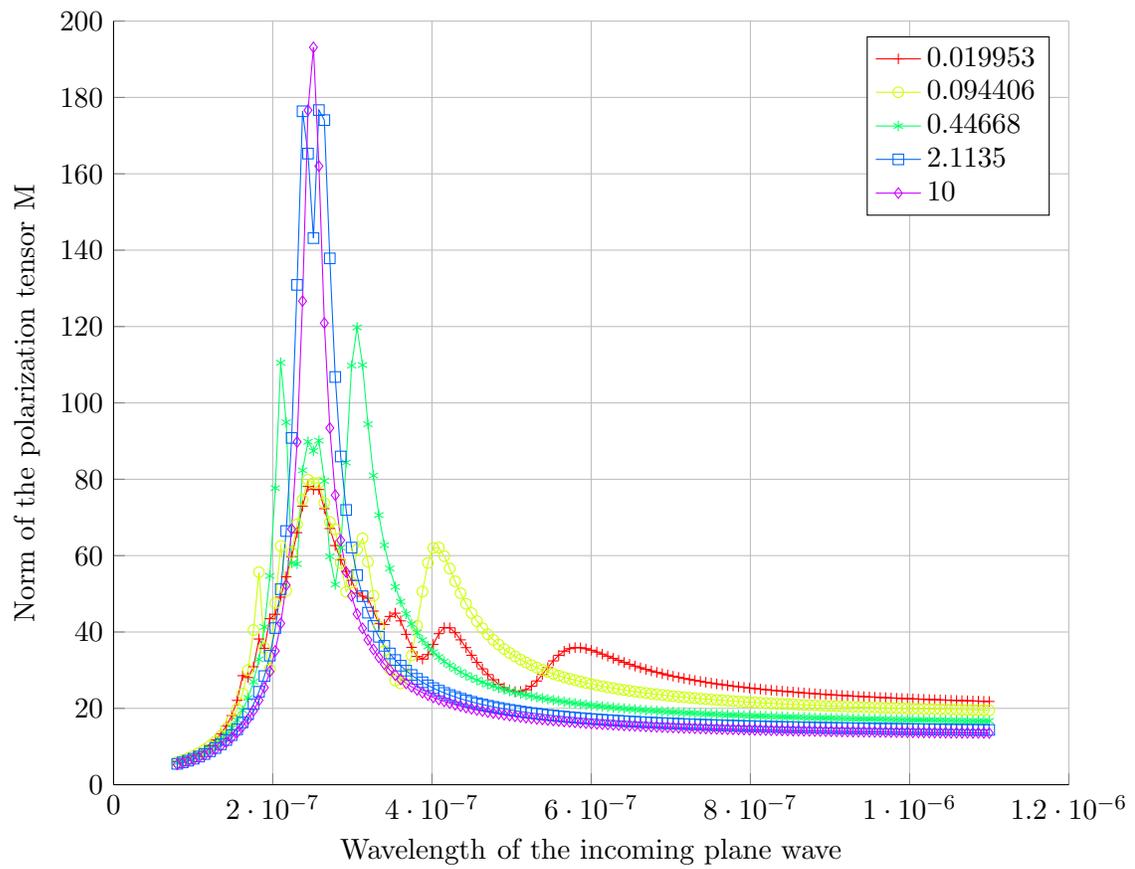}
\caption{ Norm of the polarization tensor for multiple disks for various separating distances.\label{multiplenorm} }
\end{center}
\end{figure}

Finally, it is shown in Figure \ref{multiplenorm} that when two disks are close to each other, a strong interaction occurs and the plasmonic resonance frequencies are close to those of an equivalent ellipse.  

\begin{figure}
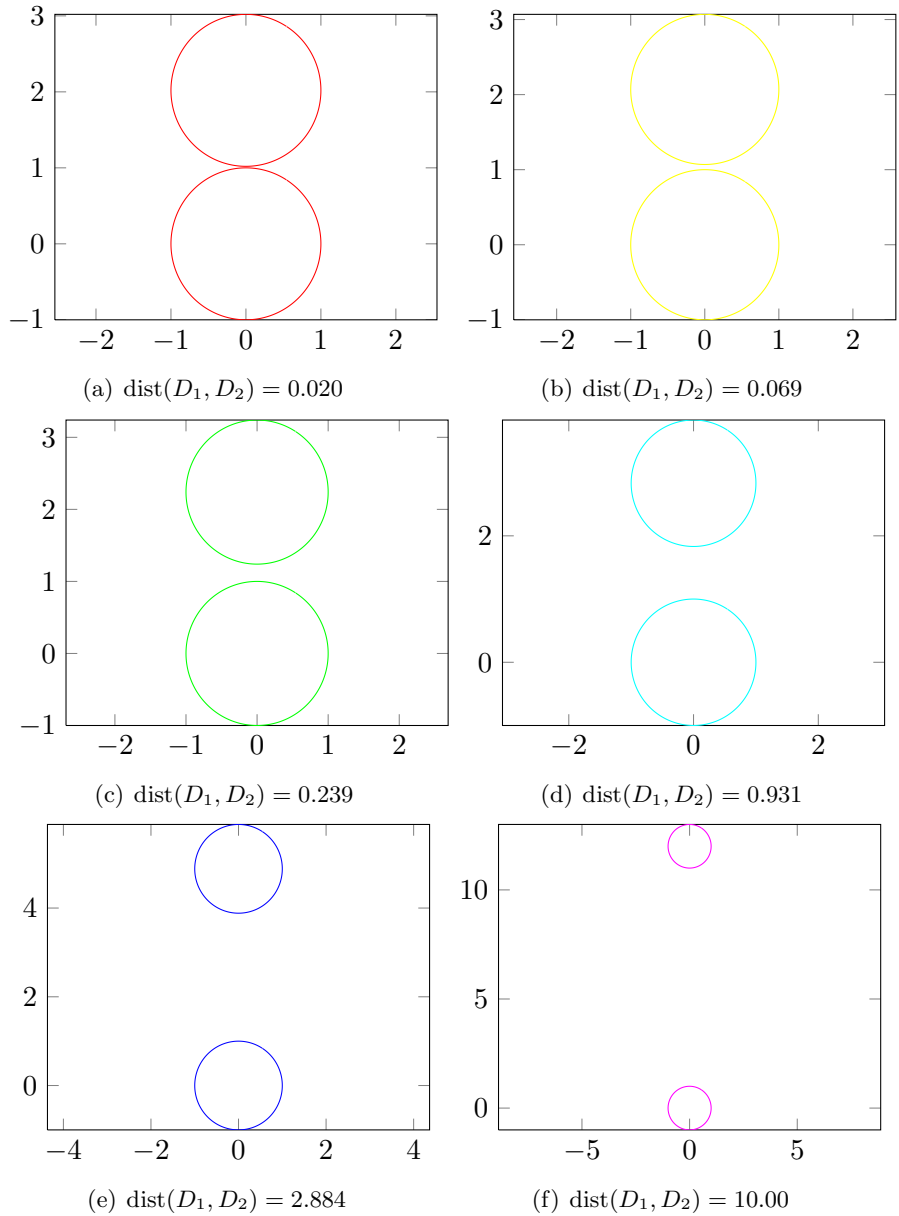

    \centering
    \subfigure[$\mathrm{dist}(D_1,D_2)=0.020$]{\label{sub1} \input{figures/polarizationcouple1.tex}}
    \subfigure[$\mathrm{dist}(D_1,D_2)=0.069$]{\label{sub2} \input{figures/polarizationcouple2.tex}}
    \subfigure[$\mathrm{dist}(D_1,D_2)=0.239$]{\label{sub3} \input{figures/polarizationcouple3.tex}}
    \subfigure[$\mathrm{dist}(D_1,D_2)=0.931$]{\label{sub4} \input{figures/polarizationcouple4.tex}}
    \subfigure[$\mathrm{dist}(D_1,D_2)=2.884$]{\label{sub5} \input{figures/polarizationcouple5.tex}}
    \subfigure[$\mathrm{dist}(D_1,D_2)=10.00$]{\label{sub6} \input{figures/polarizationcouple6.tex}}
    \caption{Different couplings between two disks.}
    \label{fig:configurations}
\end{figure}

%\begin{figure}
%\begin{center}\label{fig:polarizationnormcouple}
%\input{figures/polarizationnorm.tex}
%\caption{ Norm of the polarization tensor for a couple of spheres for various distances.}
%\end{center}
%\end{figure}

\section{Concluding remarks}

In this paper, we have provided a mathematical framework for localized plasmon resonance of nanoparticles. We have derived a uniform small volume expansion for the solution to Maxwell's equations in the presence of nanoparticles excited at their plasmonic resonances. We have presented a variety of  numerical results to illustrate our main findings.  As the particle size increases and moves away from the quasi-static approximation, high-order polarization tensors \cite{book2} should be included in order to compute the plasmonic resonances, which become size-dependent. This would be the subject of a forthcoming work. The scalar case was recently considered in \cite{matias}. Our approach in this paper, combined with the ones in \cite{2nd, iakovleva}, opens also a door for a numerical and mathematical framework for optimal shape design of resonant nanoparticles and their superresolved imaging.

%One of the most challenging problems in nanoparticle imaging is to study the heat generation in nanopaticles when illuminated at their plasmonic resonance. When a metal nanoparticle is illuminated, part of the intercepted light is scattered in the surroundings, while the other part gets absorbed and dissipated into heat.  Thermoplasmonics, which takes advantage of photothermal effects induced by resonant light absorption in metallic nanostructures, has emerged as a new research direction enabling nanoscale heat sources. These effects have been investigated for different applications, such as photothermal cancer therapy. Heat generation in plasmonic nanoparticles induced by light absorption has long been considered only as a side effect, which had to be minimized, for instance, for fluorescence enhancement. Since recently, it has been realized and demonstrated that metal nanoparticles can play the role of efficient nanosources of heat, which opens up a new emerging set of photoacoustic imaging techniques.  Modeling heat exchange between two or more nanoparticles has been the subject of intensive activity. The near-field interaction between two neighboring nanoparticles is known to enhance radiative heat transfer. 
%The understanding of this phenomena would be the subject of a forthcoming paper. Finally, it would be also interesting to apply our results in this paper to photoacoustic imaging.  
%

\appendix
\section{Jump formula} \label{jumpMproof}

We want to prove the jump formula (\ref{jumpM}) for $\nubf \times \g \times \mathcal{A}_D$. The continuity of $\mathcal{A}_D^k[\phibf]$ is a consequence of the continuity of single layer potentials. 
Assume that $\phibf$ is a continuous tangential field. We first prove the jump relation for $k=0$.
For $z\in \mathbb{R}^3\setminus \dr D$, $$\g\times \mathcal{A}_D[\phibf] (z) = \int_{\dr D} \g_z \times \left( \phibf(y) \Gamma(z,y)\right) d\sigma(y).$$
So if $x\in \dr D$ and $z=x+h\nubf(x)$, then by using vector calculus we have:
\begin{equation*}
\nubf(x) \times \g \times \mathcal{A}_D[\phibf](z)= \int_{\dr D} \big[\left(\phibf(y) \cdot \nubf(x)\right)  \g_z \Gamma(z,y) - \left(\g_z \Gamma(z,y) \cdot \nubf(x)\right) \phibf(y) \big]d\sigma(y).
\end{equation*}
Since $\phibf$ is tangential, we have $\forall y\in \dr D, \nubf(y)\cdot \phibf(y) =0$, so we can write 
\begin{equation*}
\nubf(x) \times \g \times \mathcal{A}_D[\phibf](z)= \int_{\dr D} \big[\left(\phibf(y) \cdot [\nubf(x)-\nubf(y)]\right)  \g_z \Gamma(z,y) - \left(\g_z \Gamma(z,y) \cdot \nubf(x) \right) \phibf(y) \big]d\sigma(y).
\end{equation*}
Here, following the same idea as the one in the proof of the jump of the double layer potential in \cite{colton2013integral}, we introduce
\begin{equation*}
\mathcal{D}_D[1](z)= \int_{\dr D}\frac{\dr \Gamma}{\dr \nubf (y)} (z,y) d\sigma(y), \quad z\in \mathbb{R}^3\setminus \dr D,
\end{equation*}
which takes the following values (\cite[p. 48]{colton2013integral}):
\begin{equation}
\mathcal{D}_D[1](z)= \left\{ \begin{aligned}
 & 0 \quad \mbox{ if } & z\in \mathbb{R}^3\setminus \overline{D},\\
 &-\frac{1}{2} \quad \mbox{ if } &  z\in \dr D,\\
 &-1 \quad \mbox{ if } & z\in D. 
 \end{aligned}\right.
\end{equation}
We write  $$\nubf(x) \times \g \times \mathcal{A}_D[\phibf](z) = \phibf(x) \mathcal{D}_D[1](z)  + {f}(z)$$ with
\begin{multline*}
{f}(z) = \int_{\dr D} \bigg[\left(\phibf(y) \cdot [\nubf(x)-\nubf(y)]\right)  \g_z \Gamma(z,y)\\ - \left(\g_z \Gamma(z,y) \cdot \left[\nubf(x) -\nubf(y)\right] \right) \phibf(y) - \left(\g_z \Gamma(z,y) \cdot \nubf(y) \right) \phibf(y) -\frac{\dr \Gamma}{\dr \nubf (y)} (z,y)\phibf(x)  \bigg]d\sigma(y).
\end{multline*}
Using the fact that $\g_z \Gamma(z,y) = -\g_y \Gamma (z,y)$ we get 
\begin{multline}\label{deff}
{f}(z) = \int_{\dr D} \bigg[\left(\phibf(y) \cdot [\nubf(x)-\nubf(y)]\right)  \g_z \Gamma(z,y)\\ - \left(\g_z \Gamma(z,y) \cdot \left[\nubf(x) -\nubf(y)\right] \right) \phibf(y)  +\frac{\dr \Gamma}{\dr \nubf(y)}(z,y)\left( \phibf(y) -\phibf(x)\right) \bigg]d\sigma(y).
\end{multline}
Now, we have only to prove that ${f}$ is continuous across $\dr D$, i.e., when $t\rightarrow 0$, ${f}(z)= {f}(x+t\nubf(x)) \longrightarrow {f}(x)$.
If we assume that it is true, then we can write for $z \in \mathbb{R}^3 \setminus \overline{D}$,
\begin{equation*}
\nubf(x) \times \g \times \mathcal{A}_D[\phibf](z) =\big[\phibf(x)  \mathcal{D}_D[1](z)- \phibf(x) \mathcal{D}_D[1](x) + {f}(z)\big] - \frac{\phi}{2}(x),
\end{equation*}
since $\mathcal{D}_D[1](x) = -1/2$.
So, when $t\rightarrow 0^+$, we get
\begin{equation*}
\big( \nubf(x) \times \g \times \mathcal{A}_D[\phibf](x) \big)^+=\big[ -\phibf(x) \mathcal{D}_D[1](x) + {f}(x) \big] - \frac{\phi}{2}(x).
\end{equation*}
Now we see that since $\phibf(y)\cdot \nubf(y) = 0, \quad  \forall y \in \dr D$
\begin{multline*}
-\phibf(x) \mathcal{D}_D[1](x) + {f}(x)=-\int_{\dr D}\frac{\dr \Gamma}{\dr \nubf (y)} (x,y) \phibf(x)  d\sigma(y) \\+\int_{\dr D} \bigg[\left(\phibf(y)\cdot\nubf(x)\right) \g_x\Gamma(x,y) -\left(\g_x \Gamma(x,y) \cdot \nubf(x)\right) + \frac{\dr \Gamma}{\dr \nu(y)}(x,y) \phibf(x) \bigg] d\sigma(y),
\end{multline*} which is exactly
\begin{equation*}
-\phibf(x) \mathcal{D}_D[1](x) + {f}(x) = \int_{\dr D} \nubf(x) \times \g_x \times \left[\Gamma(x,y) \phibf(y)\right] d\sigma(y).
\end{equation*}
So the limit can be expressed as 
\begin{equation*}
\big( \nubf(x) \times \g \times \mathcal{A}_D[\phibf](x) \big)^+=\int_{\dr D} \nubf(x) \times \g_x \times \left[\Gamma(x,y) \phibf(y)\right] d\sigma(y) - \frac{\phi}{2}(x).
\end{equation*}
The limit when $t\rightarrow 0^-$ is computed similarly and we find (\ref{jumpM}) for $k=0$. The extension to $k>0$ can be done because the difference between the double layer potential with kernel $\Gamma^k$ and $\Gamma$ is continuous; see, for instance, \cite[p.47]{colton2013integral}.

Now, we go back to the continuity of ${f}$ defined by (\ref{deff}). We apply several results from \cite{colton2013integral} to get the continuity.
The following lemma, which we state  here for the sake of completeness,  can be found in \cite{colton2013integral}.
\begin{lemm}\label{conti1}
 Assume that the kernel $K$ is continuous for all $x$ in a neighborhood $D_h$ of $\dr D$, $y\in \dr D$ and $x\neq y$. Assume that there exists $M>0$ such that $$\vert K(x,y) \vert \leq M \vert x-y \vert^{-2}$$ and assume that there exists $m\in \mathbb{N}$ such that $$\vert K(x_1,y)-K(x_2,y)\vert \leq M\sum_{j=1}^m \vert x_1-y\vert^{-2-j} \vert x_1 - x_2\vert^j$$ for all $x_1,x_2\in D_h$, $y\in \dr D$ with $2\vert x_1- x_2\vert \leq \vert x_1-y\vert$ and that $$\left \vert \int_{\dr D \setminus S_{x,r}} K(z,y) d\sigma(y) \right\vert \leq M$$ for all $x\in \dr D$ and $z=x+h\nubf(x) \in D_h$ and all $0<r<R$.
Then, $$u(z) = \int_{\dr D} K(z,y)[\phi(y)-\phi(x)] d\sigma(y)$$  belongs to $\mathcal{C}^{0,\alpha}(D_h)$ if $\phi \in \mathcal{C}^{0,\alpha}(\dr D)$.
\end{lemm}
It can be shown that \begin{equation*}
\left\vert \frac{\dr \Gamma(x,y)}{\dr \nubf(y)} -  \frac{\dr \Gamma(z,y)}{\dr \nubf(y)}\right\vert \leq  C \frac{\vert x-z\vert}{\vert z-y\vert^3}.
\end{equation*}
Using the above lemma with $m=1$ and the kernel associated with the double layer potential gives
 $$
\int_{\dr D}\frac{\dr \Gamma}{\dr \nubf(y)}(z,y)\left[ \phibf(y) -\phibf(x)\right]  d\sigma(y) \longrightarrow \int_{\dr D}\frac{\dr \Gamma}{\dr \nubf(y)}(x,y)\left[ \phibf(y) -\phibf(x)\right]  d\sigma(y) 
$$  as $z\rightarrow x\in \dr D$.

We now make use of the following lemma from \cite{colton2013integral}.
\begin{lemm}\label{conti2}
 Assume that the kernel $K(x,y)$ is continuous for all $x$ in a closed domain $\Omega$ containing $\dr D$ in its interior, $y\in \dr D$ and $x\neq y$. Assume that there exists $M>0$ and $\alpha \in]0,2]$ such that $$\vert K(x,y) \vert \leq M \vert x-y \vert^{\alpha-2}$$ and assume that there exists $m\in \mathbb{N}$ such that $$\vert K(x_1,y)-K(x_2,y)\vert \leq M\sum_{j=1}^m \vert x_1-y\vert^{\alpha-2-j} \vert x_1 - x_2\vert^j$$ for all $x_1,x_2\in D_h$, $y\in \dr D$ with $2\vert x_1- x_2\vert \leq \vert x_1-y\vert$.
Then $$u(x) = \int_{\dr D} K(x,y)\phi(y)d\sigma(y), \quad x\in \Omega$$  belongs to $\mathcal{C}^{0,\beta}(\Omega)$ if $\phi \in \mathcal{C}^{0,\alpha}(\dr D)$. $\beta \in ]0,\alpha]$ if $\alpha \in ]0,1[ $, $\beta \in ]0,1[$ if $\alpha =1$ and $\beta \in ]0,1] $ if $ \alpha \in]1,2[$.
\end{lemm}

Using the fact that $\dr D$ is of class $\mathcal{C}^2$, we have 
$$\vert \nubf(x) - \nubf(y) \vert \leq \vert x - y \vert, \quad  \forall x,y \in \dr D.$$ 
We can apply Lemma \ref{conti2} with $\alpha =1$ and $m=1$ to the second and third terms of
 ${f}$ and get the continuity of
\begin{equation}
\int_{\dr D} \bigg[\left(\phibf(y) \cdot [\nubf(x)-\nubf(y)]\right)  \g_z \Gamma(z,y)- \left(\g_z \Gamma(z,y) \cdot \left[\nubf(x) -\nubf(y)\right] \right) \phibf(y) \bigg]d\sigma(y)
\end{equation}
when $ z\rightarrow x\in  \dr D$, which conclude the proof for a continuous tangential field $\phi$. The formula can be extended to $L^2_T$ by a density argument .

\end{document}